\documentclass[12pt,a4paper]{article}
\usepackage{amsfonts}
\usepackage{amsmath}
\DeclareMathOperator*{\argmin}{argmin} 
\usepackage{mathtools}
\usepackage{array}
\usepackage{rotating}
\usepackage{amssymb}
\usepackage{bm}
\usepackage{geometry}
\usepackage{adjustbox}
\usepackage{changepage}
\usepackage{setspace}
\usepackage{graphicx}
\usepackage{lscape}
\usepackage{verbatim}
\usepackage{natbib}
\usepackage{calrsfs}
\usepackage{mathrsfs}
\usepackage{algorithm2e}
\usepackage{algpseudocode}
\usepackage{tablefootnote}
\usepackage{tabularx}
\usepackage{threeparttablex}
\usepackage{ragged2e}
\usepackage{booktabs}  
\usepackage[table]{xcolor}
\usepackage{longtable}
\usepackage{enumitem}
\usepackage{pdflscape}

\definecolor{winered}{rgb}{0.5,0,0}
\usepackage[bookmarks=true, bookmarksnumbered=true, allbordercolors={1 1 1}]{hyperref}
\hypersetup{
	colorlinks   = true, 
	urlcolor     = blue, 
	linkcolor    = blue, 
	citecolor    = winered,
}
\usepackage[open=true, numbered=true]{bookmark}
\usepackage{float}
\usepackage{fancyhdr}
\usepackage[toc,page]{appendix}
\usepackage{scalefnt}
\usepackage{afterpage}
\usepackage[left]{lineno}
\usepackage{caption}
\usepackage{subcaption}
\usepackage{varioref}
\usepackage{authblk}
\usepackage{cleveref}
\usepackage[colorinlistoftodos]{todonotes}
\usepackage{multirow}
\usepackage{placeins}

\DeclareFontFamily{OT1}{pzc}{}
\DeclareFontShape{OT1}{pzc}{m}{it}{<-> s * [0.900] pzcmi7t}{}
\DeclareMathAlphabet{\mathpzc}{OT1}{pzc}{m}{it}

\makeatletter
\def\blfootnote{\xdef\@thefnmark{}\@footnotetext}
\makeatother

\emergencystretch=\maxdimen
\begin{document}		
	\title{\LARGE{Forecasting Oil Prices Across the Distribution: A Quantile VAR Approach}\blfootnote{We would like to thank Christiane Baumeister, Jamie L. Cross, Howard Bondell, Sylvia Kaufmann and participants at the 2025 European Seminar on Bayesian Econometrics (ESOBE) in Melbourne, and at the 2025 Milan Time Series Seminars (MiTSS) in Milan for their valuable comments. This paper is part of the research activities at the Centre for Applied Macroeconomics, Commodity Prices (CAMP) at the BI Norwegian Business School. The usual disclaimers apply.}}
	\author[1,4]{Hilde C. Bj\o rnland}
	\author[2,3]{Nicol\'{a}s Hardy}
	\author[3,4]{Dimitris Korobilis} 
	\affil[1]{{\footnotesize BI Norwegian Business School}}
	\affil[2]{{\footnotesize Facultad de Administracion y Economia, Universidad Diego Portales}}
	\affil[3]{{\footnotesize Adam Smith Business School, University of Glasgow}}
	\affil[4]{{\footnotesize Centre for Applied Macroeconomics and Commodity Prices (CAMP)}}
	
	\date{\today}
	
	\maketitle
	\begin{abstract}
		\noindent 
		We develop a Quantile Bayesian Vector Autoregression (QBVAR) to forecast real oil prices across different quantiles of the conditional distribution. The model allows predictor effects to vary across quantiles, capturing asymmetries that standard mean-focused approaches miss. Using monthly data from 1975 to 2025, we document three findings. First, the QBVAR improves median forecasts by 2-5\% relative to Bayesian VARs, demonstrating that quantile-specific dynamics matter even for point prediction. Second, uncertainty and financial condition variables strongly predict downside risk, with left-tail forecast improvements of 10-25\% that intensify during crisis episodes. Third, right-tail forecasting remains difficult; stochastic volatility models dominate for upside risk, though forecast combinations that include the QBVAR recover these losses. The results show that modeling the conditional distribution yields substantial gains for tail risk assessment, particularly during major oil market disruptions.
		
		\bigskip
		
		
		\bigskip
		
	\end{abstract}
	
	\thispagestyle{empty} 
	\newpage
	\onehalfspace
	\setcounter{page}{1}
	
	
	\section{Introduction}

Oil prices play a central role in shaping global economic activity, yet forecasting them remains exceptionally difficult. The comprehensive survey by \citet{AlquistKilianVigfusson2013} establishes that oil futures prices fail to systematically outperform a simple no-change forecast at horizons up to one year, while \citet{EllwangerSnudden2023} show that even sophisticated models struggle against a strengthened random walk benchmark\footnote{\citet{EllwangerSnudden2023} recommend using end-of-month daily prices rather than monthly averages to construct the random walk benchmark. Since real oil prices require deflation by the monthly CPI, we cannot implement this refinement. Nevertheless, monthly average real prices remain a relevant forecasting target for macroeconomic analysis and policy applications, where oil prices enter inflation forecasts and feed into planning decisions based on average costs over billing or budgeting cycles.}. Crisis periods expose complete forecast breakdown, cf.\ \citet{BaumeisterKilian2016}: neither futures markets nor econometric models anticipated the 2014--2016 collapse, the April 2020 negative price event, or the 2022 Russia-Ukraine spike. Model combination approaches have improved average forecasts \citep[cf.][]{BaumeisterKilianLee2014,BaumeisterKilian2015,AagBjornlandEliassen2024}, but offer limited insight into how predictive relationships differ across the distribution of oil price changes and can perform poorly during unusually large price movements. This limitation reflects a deeper challenge: oil price changes exhibit a heavy-tailed and asymmetric distribution, and standard econometric approaches, including Bayesian VARs with stochastic volatility \citep{CarrieroClarkMarcellino2016,BaumeisterKorobilisLee2022} or time-varying parameters, summarize uncertainty around a single conditional mean and therefore provide limited insight into the likelihood or nature of tail events.

\begin{figure}[t!]
\centering
\includegraphics[width=0.9\textwidth, trim={0 0 0 0}, clip]{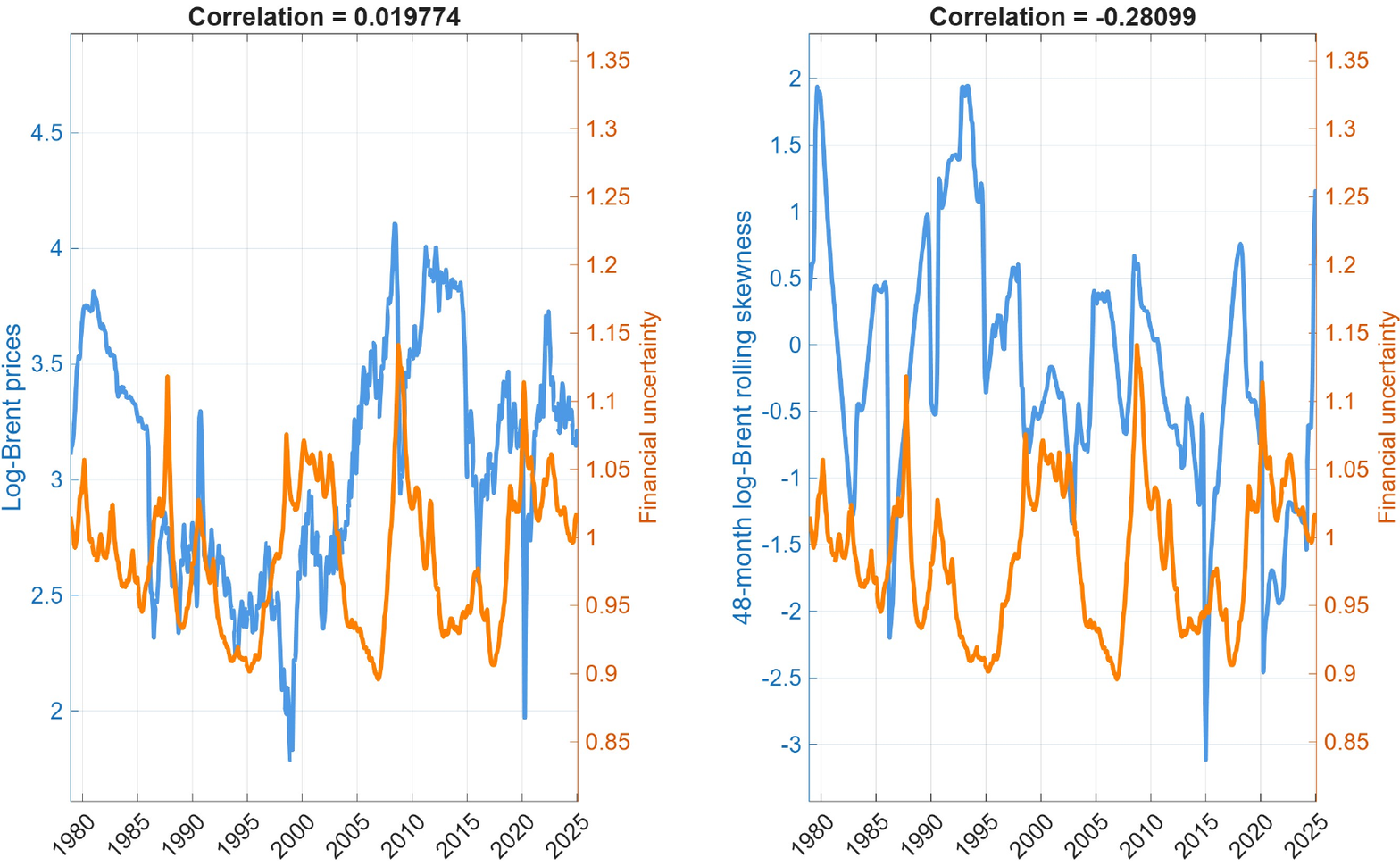}
\caption{\textbf{Brent oil prices, financial uncertainty, and distributional asymmetry:} \\ 
The left panel shows the evolution of log Brent oil prices and financial uncertainty over time. The two series are effectively uncorrelated, with a correlation coefficient of 0.02. 
The right panel plots financial uncertainty against a time-varying measure of skewness in log Brent prices. Financial uncertainty is negatively correlated with the third moment of log Brent prices ($=-0.281$). 
Time-varying skewness is computed using sample skewness over 48-month rolling windows. Financial uncertainty data are from \cite{jurado2015measuring}.}
\label{fig:correlations}
\end{figure}

Figure~\ref{fig:correlations} illustrates why mean-based approaches may miss economically relevant variation. Financial uncertainty is essentially orthogonal to the level of real Brent prices (left panel), yet systematically associated with time variation in return skewness (right panel). Periods of elevated uncertainty coincide with pronounced shifts in distributional asymmetry, even when the conditional mean exhibits little response.\footnote{This pattern is robust to alternative rolling windows, oil price measures, and transformations, and similar relationships obtain for other predictors such as geopolitical risk and financial spreads.} This suggests that macro-financial predictors may primarily load on the tails of the conditional distribution rather than on its center, motivating a forecasting framework that explicitly models heterogeneity across quantiles.

We address these challenges by developing a Quantile Bayesian Vector Autoregression (QBVAR) that models the entire conditional distribution of real oil price movements, allowing the dynamics and predictive content of economic fundamentals to vary across quantiles. Drawing on the Growth-at-Risk literature \citep{AdrianBoyarchenkoGiannone2019}, which shows that financial conditions affect the conditional distribution of macroeconomic outcomes asymmetrically, we extend this insight to oil markets, where similar asymmetries may arise from the distinct mechanisms driving price collapses versus spikes. The distributional perspective enables us to identify which variables help predict downside and upside risks, evaluate forecast accuracy specifically in the tails, and assess performance during major episodes of oil market stress.

Our methodological contribution builds on two literatures. First, the Structural VAR tradition that dominates oil market analysis, tracing back to \citet{Hamilton1983} and developed through frameworks that treat oil prices as endogenous,\footnote{For earlier SVAR studies separating oil price and macroeconomic shocks, see \citet{BurbidgeHarrison1984}, \citet{GisserGoodwin1986}, \citet{BernankeGertlerWatson1997}, and \citet{Bjornland2000}.} including \citet{Kilian2009} and the Bayesian SVAR of \citet{BaumeisterHamilton2019}. These methods have transformed our understanding of oil--macroeconomy linkages but focus on conditional means. Second, the Bayesian quantile regression literature, where \citet{YuMoyeed2001} established that maximizing the asymmetric Laplace likelihood yields valid quantile inference, and \citet{KozumiKobayashi2011} developed the Gibbs sampling algorithms that make Bayesian QVAR computationally feasible. Crucially, \citet{Srirametal2013} show that posterior inference concentrates around true quantile coefficients even when the asymmetric Laplace distribution is misspecified -- a ``working likelihood'' result that justifies our parametric approach when the true error distribution is unknown.

Our approach relates to, but is distinct from, recent work on oil price uncertainty and tail risks. \citet{AastveitCrossVanDijk2023} construct time-varying predictive densities from mixtures of conditional means and variances but do not model quantile-specific dynamics explicitly. \citet{Baumeisteretal2024} employ Bayesian Additive Regression Trees to generate nonlinear predictive densities and outperform linear quantile regressions, but derive tail forecasts from the predictive distribution rather than modeling quantile-specific dynamics directly. \citet{CarrieroClarkMarcellino2024} show that Bayesian VARs with stochastic volatility capture macroeconomic tail risks comparably to quantile regression, but symmetric distributional assumptions preclude predictor effects that vary across quantiles. By contrast, our Bayesian QVAR models the conditional distribution directly through quantile-specific coefficients, shocks, and factor structures within a multivariate framework. In addition, our factor-based specification avoids the variable-ordering dependence inherent in Cholesky-based QVAR implementations \citep{ChavleishviliManganelli2024}, a concern shown by \citet{AriasRubioShin2023} to substantially affect VAR forecasting performance. The Bayesian framework further enables flexible regularization through hierarchical shrinkage priors, which is particularly valuable in the oil forecasting context where VARs with 12 or more lags are common.

Using monthly data from 1975 to 2025, we assess the out-of-sample performance of the QBVAR across horizons and quantiles, showing substantial gains relative to standard Bayesian VARs, Bayesian VARs with stochastic volatility, and the random walk.\footnote{In unreported analyses, we also considered a BVAR with factor stochastic volatility and leverage effects, following \cite{KastnerFruhwirthSchnatterLopes2017}. This specification generally underperforms the BVAR-SV without leverage across most quantiles and horizons, so we retain the BVAR-SV as our primary stochastic volatility benchmark. Results can be obtained at request.} Three stylized facts emerge. First, the QBVAR improves median forecasts uniformly, achieving 2--5\% gains over standard BVARs -- noteworthy given the difficulty of improving upon well-specified models for point prediction. Second, uncertainty and financial condition variables strongly predict downside risk, with left-tail improvements of 10--25\% that intensify during crisis episodes, echoing the Growth-at-Risk finding that financial conditions affect the left tail more strongly than the right. Third, right-tail forecasting remains challenging; the QBVAR typically underperforms stochastic volatility models for upside risk, though forecast combinations recover this deficit. These results underscore the importance of modelling distributional dynamics when loss functions are asymmetric: variables with limited value for mean forecasts, such as crack spreads, financial uncertainty, or geopolitical risk, contain substantial information for the tails, with uncertainty and financial stress predicting downside risk through precautionary demand channels (e.g., \citealp{Bloom2009,BaumeisterKilian2016}) and demand-side measures providing some information about upside risk.

This paper makes three contributions. First, we develop a Bayesian Quantile VAR that models the entire conditional distribution of oil price movements through quantile-specific parameters and factor structures, offering an ordering-invariant alternative to existing implementations. Second, we provide new evidence that the determinants of oil price movements differ markedly across quantiles, revealing asymmetric mechanisms hidden in mean-based or density-forecasting models. Third, we show that the QBVAR delivers substantial tail-forecasting gains relative to both random-walk and state-of-the-art BVAR benchmarks, especially during major oil market disruptions. Together, these contributions demonstrate that modelling the conditional distribution, rather than only the mean or an implied density, yields meaningful improvements for risk assessment and policy-relevant forecasting.

The remainder of the paper is organized as follows. Section 2 outlines the Bayesian QVAR methodology. Section 3 describes the data, forecasting design, and out-of-sample results. Section 4 investigates forecast combination strategies and evaluates performance during major oil price episodes. Section 5 concludes.

	
\section{Econometric methodology} \label{sec:Econometrics}
	
\subsection{Quantile vector autoregression with factor structure}
	
	Standard Bayesian VARs model the conditional mean of economic variables and characterize uncertainty through parametric distributional assumptions. While effective for point forecasting, this approach provides limited insight into tail risks or distributional asymmetries that may be crucial for policy and risk management. We develop a Quantile Vector Autoregression (QVAR) that directly models specific quantiles of the conditional distribution, allowing the dynamics and predictor effects to vary across different points of the distribution.
	
	Our starting point is the Bayesian VAR with factor decomposition in the covariance matrix, as specified by \cite{Korobilis2022}. For an $n \times 1$ vector of endogenous variables $\bm y_{t}$, this takes the form
	\begin{equation}
		\bm y_{t}  = \bm \Phi \bm x_{t} + \bm \Lambda \bm f_{t} + \bm v_{t},  \quad \bm v_{t} \sim N \left(\bm 0, \bm \Sigma \right), \label{VAR}
	\end{equation}
	where $\bm x_{t} = \left(\bm 1, \bm y_{t-1},..., \bm y_{t-p} \right)'$ contains an intercept and $p$ lags, $\bm \Phi$ is an $n \times (np+1)$ coefficient matrix, $\bm f_{t} \sim N\left( \bm 0, \bm I_r \right)$ is an $r \times 1$ vector of latent factors with $r \ll n$, $\bm \Lambda$ is an $n \times r$ matrix of factor loadings, and $\bm \Sigma = \text{diag}(\sigma_{1}^{2},...,\sigma_{n}^{2})$ is diagonal. This specification implies that cross-equation comovements are captured entirely by the common component $\bm \Lambda \bm f_{t}$, with the VAR covariance matrix given by $\text{cov} \left( \bm y_{t} \vert \bm x_{t} \right) = \bm \Lambda \bm \Lambda' + \bm \Sigma$.
	
	We extend this framework to model specific quantiles of the conditional distribution. For quantile level $q \in (0,1)$, our Quantile BVAR (QBVAR) takes the form
	\begin{equation}
		\bm y_{t} = \bm \Phi_{q} \bm x_{t} + \bm \Lambda_{q} \bm f_{t,q} + \bm v_{t,q}, \quad \bm v_{t,q} \sim \prod_{i=1}^{n} \text{AL} \left(0, \sigma_{i,q}, q \right), \label{QVAR}
	\end{equation}
	where $\text{AL}(0,\sigma_{i,q},q)$ denotes the asymmetric Laplace distribution with location zero, scale $\sigma_{i,q}$, and asymmetry parameter $q$. Unlike the mean VAR in \eqref{VAR}, the QVAR features quantile-specific coefficients $\bm \Phi_{q}$, loadings $\bm \Lambda_{q}$, and factors $\bm f_{t,q}$, following the quantile factor approach of \cite{KorobilisSchroeder2024a}. This allows the model to capture different factor structures at different points of the distribution---the factors driving median dynamics may differ substantially from those driving tail behavior. We normalize $\bm f_{t,q} \sim N \left( \bm 0, \bm I_r \right)$ for all $q$.
	
	\cite{ChavleishviliManganelli2024} develop a frequentist QVAR that models quantile dynamics through a recursive triangular structure, relying on Cholesky decomposition to orthogonalize the system. While computationally convenient, this approach introduces variable ordering dependence. \cite{AriasRubioShin2023} demonstrate that such ordering choices can substantially affect forecasting performance in VARs with stochastic volatility, a concern that extends naturally to quantile settings where the conditional distribution varies across quantiles.
	
	Our factor-based specification avoids this limitation. The common component $\bm \Lambda_{q} \bm f_{t,q}$ captures cross-equation comovements without requiring identification restrictions on the loadings, yielding forecasts that are invariant to variable ordering. This is particularly valuable for forecasting applications where the goal is distributional characterization rather than structural identification.
	
	\subsection{Bayesian estimation}
	
	To facilitate posterior computation, we exploit the location-scale mixture representation of the asymmetric Laplace distribution \citep{KozumiKobayashi2011}. The QVAR likelihood in \eqref{QVAR} can be equivalently written as
	\begin{equation}
		\bm y_{t} = \bm \Phi_{q} \bm x_{t} + \bm \Lambda_{q} \bm f_{t,q} + \theta(q) \bm z_{t,q} + \bm \epsilon_{t,q}, \quad \bm \epsilon_{t,q} \sim N_{n} \left( \bm 0, \tau^{2}(q) \, \text{diag}(\bm z_{t,q}) \bm \Sigma_{q}  \right), \label{QVAR_mixture}
	\end{equation}
	where $\bm z_{t,q} = (z_{1t,q}, \ldots, z_{nt,q})'$ with $z_{it,q} \stackrel{iid}{\sim} \text{Exp}(1)$, and the quantile-specific constants are $\theta(q) = \frac{1-2q}{q(1-q)}$ and $\tau^{2}(q) = \frac{2}{q(1-q)}$. The term $\theta(q) \bm z_{t,q}$ generates asymmetry: it equals zero at the median ($q=0.5$) but shifts the conditional mean for other quantiles, with the direction depending on whether $q$ is below or above 0.5.
	
	This mixture representation renders the likelihood conditionally Gaussian in $(\bm \Phi_q, \bm \Lambda_q, \bm f_{t,q})$ given the auxiliary variables $\bm z_{t,q}$, enabling straightforward Gibbs sampling.
	
	We employ hierarchical shrinkage priors that adapt to the sparsity structure of the data. For the VAR coefficients, we use a horseshoe prior \citep{CarvalhoPolsonScott2010}:
	\begin{equation}
		\phi_{ij,q} \sim N(0, \psi_{ij,q}^2 \kappa_q^2), \quad \psi_{ij,q} \sim C^{+}(0,1), \quad \kappa_q \sim C^{+}(0,1),
	\end{equation}
	where $\phi_{ij,q}$ denotes the $(i,j)$th element of $\bm \Phi_q$, $\psi_{ij,q}$ are local shrinkage parameters, $\kappa_q$ is the global shrinkage parameter, and $C^{+}(0,1)$ denotes the half-Cauchy distribution. This prior aggressively shrinks irrelevant coefficients toward zero while allowing important predictors to remain unshrunk.
	
	For the factor loadings, we use independent normal priors: $\lambda_{ij,q} \sim N(0, \underline{V}_\lambda)$. The idiosyncratic scale parameters receive inverse-gamma priors: $\sigma_{i,q} \sim \text{IG}(\underline{a}_\sigma, \underline{b}_\sigma)$. We set $\underline{V}_\lambda = 1$, $\underline{a}_\sigma = 3$, and $\underline{b}_\sigma = 1$ throughout.
	
	Posterior inference proceeds via Markov Chain Monte Carlo. Given the mixture representation \eqref{QVAR_mixture}, all full conditional distributions are available in closed form. Let $k = np + 1$ denote the number of regressors per equation.
	
	\begin{enumerate}
		\item[] \textit{Step 1: VAR coefficients $\bm \Phi_q$.} Define the adjusted dependent variable $\tilde{y}_{it} = y_{it} - \bm \lambda_{i,q}' \bm f_{t,q} - \theta(q) z_{it,q}$ for each equation $i = 1,\ldots,n$. The conditional posterior for each row $\bm \phi_{i,q}' = (\phi_{i1,q}, \ldots, \phi_{ik,q})$ of $\bm \Phi_q$ is
		\begin{equation}
			\bm \phi_{i,q} \mid \cdot \sim N\left( \bar{\bm \phi}_{i,q}, \bar{\bm V}_{\phi,i} \right),
		\end{equation}
		where $\bar{\bm V}_{\phi,i} = \left( \bm X' \bm D_{i,q}^{-1} \bm X + \bm V_{\phi,i}^{-1} \right)^{-1}$, $\bar{\bm \phi}_{i,q} = \bar{\bm V}_{\phi,i} \bm X' \bm D_{i,q}^{-1} \tilde{\bm y}_i$, with $\bm X = (\bm x_1, \ldots, \bm x_T)'$, $\bm D_{i,q} = \tau^2(q) \sigma_{i,q} \text{diag}(z_{i1,q}, \ldots, z_{iT,q})$, and $\bm V_{\phi,i} = \text{diag}(\psi_{i1,q}^2 \kappa_q^2, \ldots, \psi_{ik,q}^2 \kappa_q^2)$.
		
		\item[] \textit{Step 2: Factor loadings $\bm \Lambda_q$.} For each equation $i$, define $\tilde{y}_{it}^{\Lambda} = y_{it} - \bm \phi_{i,q}' \bm x_t - \theta(q) z_{it,q}$. The conditional posterior for row $\bm \lambda_{i,q}' = (\lambda_{i1,q}, \ldots, \lambda_{ir,q})$ of $\bm \Lambda_q$ is
		\begin{equation}
			\bm \lambda_{i,q} \mid \cdot \sim N\left( \bar{\bm \lambda}_{i,q}, \bar{\bm V}_{\lambda,i} \right),
		\end{equation}
		where $\bar{\bm V}_{\lambda,i} = \left( \bm F_q' \bm D_{i,q}^{-1} \bm F_q + \underline{V}_\lambda^{-1} \bm I_r \right)^{-1}$, $\bar{\bm \lambda}_{i,q} = \bar{\bm V}_{\lambda,i} \bm F_q' \bm D_{i,q}^{-1} \tilde{\bm y}_i^{\Lambda}$, and $\bm F_q = (\bm f_{1,q}, \ldots, \bm f_{T,q})'$.
		
		\item[] \textit{Step 3: Latent factors $\bm f_{t,q}$.} Define $\tilde{\bm y}_{t}^{f} = \bm y_{t} - \bm \Phi_q \bm x_t - \theta(q) \bm z_{t,q}$. The conditional posterior for each $\bm f_{t,q}$ is
		\begin{equation}
			\bm f_{t,q} \mid \cdot \sim N\left( \bar{\bm f}_{t,q}, \bar{\bm V}_{f,t} \right),
		\end{equation}
		where $\bar{\bm V}_{f,t} = \left( \bm \Lambda_q' \bm D_{t,q}^{-1} \bm \Lambda_q + \bm I_r \right)^{-1}$, $\bar{\bm f}_{t,q} = \bar{\bm V}_{f,t} \bm \Lambda_q' \bm D_{t,q}^{-1} \tilde{\bm y}_{t}^{f}$, and $\bm D_{t,q} = \tau^2(q) \text{diag}(\sigma_{1,q} z_{1t,q}, \ldots, \sigma_{n,q} z_{nt,q})$.
		
		\item[] \textit{Step 4: Auxiliary variables $\bm z_{t,q}$.} Define the residual $e_{it,q} = y_{it} - \bm \phi_{i,q}' \bm x_t - \bm \lambda_{i,q}' \bm f_{t,q}$. Each $z_{it,q}$ is drawn independently from a generalized inverse Gaussian distribution:
		\begin{equation}
			z_{it,q} \mid \cdot \sim \text{GIG}\left( \frac{1}{2}, \frac{e_{it,q}^2}{\tau^2(q) \sigma_{i,q}}, \frac{\theta(q)^2}{\tau^2(q) \sigma_{i,q}} + 2 \right),
		\end{equation}
		where $\text{GIG}(p, a, b)$ denotes the generalized inverse Gaussian with density $f(x) \propto x^{p-1} \exp\left( -\frac{1}{2}(a x^{-1} + b x) \right)$.
		
		\item[] \textit{Step 5: Scale parameters $\sigma_{i,q}$.} The conditional posterior is inverse-gamma:
		\begin{equation}
			\sigma_{i,q} \mid \cdot \sim \text{IG}\left( \underline{a}_\sigma + T, \, \underline{b}_\sigma + \sum_{t=1}^{T} \frac{e_{it,q}^2}{2 \tau^2(q) z_{it,q}} + \sum_{t=1}^{T} \frac{\theta(q)^2 z_{it,q}}{2 \tau^2(q)} \right).
		\end{equation}
		
		\item[] \textit{Step 6: Shrinkage parameters.} The horseshoe local and global shrinkage parameters $\{\psi_{ij,q}, \kappa_q\}$ are updated using the slice sampler of \cite{MakalicSchmidt2016}.
	\end{enumerate}
	
	\subsection{Quantile forecasting}
	
	A distinctive feature of quantile regression is that the asymmetric Laplace distribution serves as a computational device for targeting specific quantiles rather than a statement about the true data generating process \citep{YuMoyeed2001, Srirametal2013}. The posterior concentrates around the true quantile regression coefficients regardless of the actual error distribution, justifying the ``working likelihood'' interpretation.
	
	For multi-step forecasting, we must decide how to generate predictions beyond the one-step-ahead horizon. Recall from equation \eqref{QVAR_mixture} that the ALD mixture representation introduces auxiliary variables $\bm z_{t,q}$ that enter both the conditional mean (through $\theta(q) \bm z_{t,q}$) and the conditional variance (through $\tau^2(q) \text{diag}(\bm z_{t,q})$). Iterating the QVAR forward using this full representation would require simulating future paths of $\bm z_{t+1,q}, \bm z_{t+2,q}, \ldots$, which creates two problems. First, the exponentially-distributed auxiliary variables can generate explosive forecast paths, particularly at extreme quantiles where $\theta(q)$ is large. Second, tracking these auxiliary variables through the forecasting recursion adds computational burden without clear accuracy gains---the quantile-specific information is already encoded in the estimated coefficients $\bm \Phi_q$ and $\bm \Lambda_q$.
	
	We therefore adopt a standard VAR forecasting approach with Gaussian innovations after estimation. This is justified on two grounds. First, the ALD serves as a working likelihood for estimation: the goal is to obtain consistent estimates of the quantile-specific coefficients, not to characterize the true error distribution \citep{Srirametal2013}. Once $\bm \Phi_q$ and $\bm \Lambda_q$ are estimated, forecasts can be generated using any reasonable innovation distribution. Second, for iterated multi-step forecasts, the accumulated shocks over multiple periods converge toward Gaussianity by the central limit theorem, making the normal approximation increasingly appropriate as the forecast horizon grows.

	At each forecast origin $t$, we generate $h$-step-ahead quantile forecasts as follows. For each posterior draw $s = 1, \ldots, S$ of the parameters $\{\bm \Phi_q^{(s)}, \bm \Lambda_q^{(s)}, \bm \Sigma_q^{(s)}\}$:
	\begin{enumerate}[leftmargin=*,itemsep=2pt]
		\item Initialize $\hat{\bm y}_{t|t}^{(s)} = \bm y_t$ (the observed value at the forecast origin).
		\item For $j = 1, \ldots, h$:
		\begin{enumerate}
			\item Draw future factors: $\bm f_{t+j}^{(s)} \sim N(\bm 0, \bm I_r)$.
			\item Draw future shocks: $\bm \epsilon_{t+j}^{(s)} \sim N(\bm 0, \bm \Omega_q^{(s)})$, where $\bm \Omega_q^{(s)} = \bm \Lambda_q^{(s)} (\bm \Lambda_q^{(s)})' + \bm \Sigma_q^{(s)}$.
			\item Compute: $\hat{\bm y}_{t+j|t}^{(s)} = \bm \Phi_q^{(s)} \hat{\bm x}_{t+j|t}^{(s)} + \bm \epsilon_{t+j}^{(s)}$.
		\end{enumerate}
		\item The quantile forecast is the median across posterior draws: $\hat{Q}_t(q,h) = \text{median}_s \left\{ \hat{y}_{t+h|t}^{(s)} \right\}$.
	\end{enumerate}
	
	This procedure integrates over parameter uncertainty while using the estimated quantile-specific dynamics to generate distributional forecasts. By estimating separate QVAR models at multiple quantile levels ($q \in \{0.1, 0.5, 0.9\}$), we trace out the conditional distribution of oil prices, capturing both central tendency and tail risks. Our objective is to characterize the shape of this distribution (its location, spread, and asymmetry) rather than to quantify uncertainty around individual quantile estimates.
	
	We evaluate quantile forecasts using the quantile score (also known as the pinball loss or check function), which is the proper scoring rule for quantile forecasts \citep{GneitingRaftery2007}. For quantile level $q$ and forecast error $u_t = y_{t+h} - \hat{Q}_t(q,h)$, the quantile score is
	\begin{equation}
		\text{QS}(q) = \frac{1}{P} \sum_{t=1}^{P} \rho_{q}(u_t), \quad \text{where} \quad \rho_{q}(u) = u \cdot (q - \mathbb{I}\{u < 0\}).
	\end{equation}
	This loss function penalizes errors asymmetrically: at $q = 0.1$, overpredictions (predicting values too high) are penalized nine times more heavily than underpredictions, reflecting the greater concern about missing downside risks when forecasting the left tail. Lower quantile scores indicate more accurate forecasts.
	
	
	\section{Forecasting oil prices} \label{sec:Forecasting}
	
	\subsection{Data and forecasting design}
	
	Our empirical analysis employs monthly data spanning from January 1975 to February 2025, encompassing 602 observations and providing one of the longest time series available for oil price forecasting research. We construct two primary oil price series, both expressed in real terms using the U.S. Consumer Price Index: the Refiner Acquisition Cost (RAC) of imported crude oil from the Energy Information Administration's Monthly Energy Review, and Brent crude oil spot prices extended backward using RAC growth rates to ensure consistency with historical price movements. Both series undergo logarithmic transformation to achieve stationarity and facilitate interpretation of results in terms of percentage changes. The analysis below presents results for RAC forecasts; comparable results using Brent can be found in the online supplement.
	
	The forecasting models incorporate a comprehensive set of predictor variables capturing different dimensions of oil market fundamentals and broader economic conditions.  
These include global economic activity measures, supply-side indicators, inventory and storage data, consumption patterns, geopolitical risk indices, financial market conditions, product market spreads, and uncertainty measures. All variables undergo appropriate transformations to ensure stationarity, with detailed descriptions, sources, and transformation codes provided in the Data Appendix. This rich set of predictors allows us to evaluate which economic fundamentals are most informative for quantile-specific oil price forecasts.

The baseline specification is a three-variable VAR system comprising oil prices, a measure of global oil market fundamentals (Global Liquid Fuels Consumption), and one additional predictor variable selected from the comprehensive set described in Table~\ref{tab:data_description}. This design allows systematic evaluation of individual predictor contributions to distributional forecasts. 
Motivated by the distributional-risk perspective of the Growth-at-Risk literature
\citep{AdrianBoyarchenkoGiannone2019}, and drawing on recent oil market research for predictor selection
\citep{Baumeisteretal2024}, we focus on twelve representative predictors spanning distinct
economic categories:
\begin{itemize}[leftmargin=*, itemsep=2pt]
	\item \textbf{Uncertainty measures}: JLN (Jurado, Ludvigson, Ng macro uncertainty index) and JLNF12 (12-month ahead financial uncertainty). These capture the precautionary demand channel emphasized in the oil market literature.
	\item \textbf{Financial conditions}: EBP (Excess Bond Premium), GZ Spread (Gilchrist-Zakra\v{s}ek credit spread), and Crack spread (gasoline-crude differential). The Growth-at-Risk literature finds financial conditions strongly predict downside risk; crack spreads provide additional tail-risk information.
	\item \textbf{Global activity}: GECON (global economic conditions index) and Kilian Index (global real economic activity based on shipping costs). Standard demand-side predictors.
	\item \textbf{Supply and inventory}: OECD stocks (OECD crude oil inventories) and World Rig Count (Baker Hughes drilling activity). Capture inventory dynamics, storage arbitrage, and forward-looking production capacity.
	\item \textbf{Geopolitical risk}: GPRH (Caldara-Iacoviello Geopolitical Risk index), GPRHA (Geopolitical Acts index), and GPRHT (Geopolitical Threats index). Capture supply disruption risk and upside price pressures.
\end{itemize}

For the QBVAR, we consider lag lengths $p \in \{1, 2, 4, 12\}$. Our baseline results use $p = 4$, which balances flexibility with parameter parsimony. This is an important consideration for quantile regression, which requires sufficient observations to estimate quantile-specific coefficients reliably. The three-variable QBVAR(12) specification involves substantially more parameters, and while results remain reasonable, more parsimonious specifications have an edge on average. Results for alternative lag lengths are available in the Online Supplement. All Bayesian VAR models (QBVAR, BVAR, and BVAR-SV) employ a factor structure in the error covariance matrix with $r=1$ common factor throughout all exercises.\footnote{In exploratory analyses not reported here, we found that specifications with $r=2$ factors yield qualitatively similar results.}

We implement a recursive forecasting scheme where models are re-estimated at each forecast origin with an expanding estimation window. We generate forecasts at horizons $h = 1, 2, \ldots, 12$ months ahead for log growth rates of oil prices. To assess robustness and performance across different economic environments, we report results for two evaluation windows: 2008M1--2025M2 (205 monthly forecast origins, encompassing the financial crisis, commodity super-cycle collapse, and COVID-19 pandemic) and 2013M1--2025M2 (145 monthly forecast origins, focusing on the post-crisis period including the 2014--2016 oil price collapse). Results for additional evaluation windows are available in the Online Supplement.

We compare the QBVAR against three benchmark models representing the current state-of-the-art in oil price forecasting. First, BVAR(12), a three-variable Bayesian VAR that includes the same endogenous variables as the corresponding QBVAR specification (oil prices, Global Liquid Fuels Consumption, and the same additional predictor). To ensure comparability, both BVAR and BVAR-SV use the same factor structure in the error covariance matrix and the same horseshoe shrinkage prior on the VAR coefficients as the QBVAR. The only difference is that the BVAR targets the conditional mean with constant volatility, while the BVAR-SV incorporates stochastic volatility. Second, a no-change random walk (RW), the challenging univariate benchmark that has proven difficult to beat \citep{AlquistKilianVigfusson2013}. Third, BVAR-SV(12), a three-variable specification (again with identical endogenous variables) that incorporates stochastic volatility and represents the state-of-the-art for density forecasting \citep{CarrieroClarkMarcellino2024}. For BVAR and BVAR-SV benchmarks, we extract quantile forecasts from the posterior predictive distribution; for the QBVAR, we report the median across posterior draws from the quantile-specific model. 

Performance is assessed using quantile score (QS) ratios, with values below 1.00 indicating QBVAR outperformance. We report results separately for QS10 (left tail), QS50 (median), and QS90 (right tail) to evaluate performance across the distribution. For all Bayesian models, posterior inference is based on Markov Chain Monte Carlo simulation. At each forecast origin, we run the Gibbs sampler for 3,000 iterations, discarding the first 1,000 as burn-in. From the remaining 2,000 post-burn-in draws, we retain every 5th draw to reduce autocorrelation, yielding 400 saved posterior draws per forecast step. 

The Online Supplement provides comprehensive robustness analysis for all results presented in this paper. Specifically, it reports quantile score ratios for: (i) both oil price measures (U.S.\ Refiners' Acquisition Cost and Brent); (ii) alternative lag specifications for the QBVAR ($p \in \{1, 2, 4, 12\}$); (iii) four evaluation windows spanning different market regimes (2018M1--2025M2, 2013M1--2025M2, 2008M1--2025M2, and 2000M1--2025M2); and (iv) the complete set of twelve predictors, grouped by economic channel---uncertainty and financial conditions (JLN, JLNF12, EBP, GZ spread), global activity and supply (GECON, Kilian Index, OECD stocks, World Rig Count), and geopolitical risk (GPRH, GPRHT, GPRHA, Crack spread). These results are reported against all three benchmark models: BVAR(12), the random walk, and BVAR-SV(12). The supplement also includes forecast combination results examining optimal weighting between QBVAR and BVAR-SV forecasts.

\subsection{Comparison against BVAR(12)}

Table~\ref{tab:1} reports out-of-sample Quantile Score (QS) ratios from our QBVAR(4) model relative to the BVAR(12) benchmark, using four representative predictors that capture distinct economic channels: macroeconomic uncertainty (JLN), geopolitical risk (GPRHT), financial conditions (GZ spread), and global real activity (GECON).\footnote{The Online Supplement reports results for all twelve predictors, four different estimation samples, and QBVAR specifications with lag orders $p\in \{1,2,4,12\}$; the conclusions are qualitatively unchanged.} To confirm robustness across time periods, we report results for two distinct evaluation windows.

Two key features of these results merit discussion. First, we observe consistent improvements in the median forecast (QS50) across all predictors, with ratios below unity in nearly every case. Gains are modest but remarkably stable: the GECON Index delivers 4--5\% improvements at medium horizons, while other predictors yield 2--5\% gains that persist across both evaluation periods.

\begin{table}[t!] 
	\centering
	\small
	\caption{Out-of-sample quantile score ratios relative to BVAR(12)}
	\label{tab:1}
	\begin{tabular}{@{}c ccc ccc ccc ccc@{}}
		\toprule
		& \multicolumn{3}{c}{JLN} & \multicolumn{3}{c}{GPRHT} & \multicolumn{3}{c}{GZ spread} & \multicolumn{3}{c}{GECON} \\
		\cmidrule(lr){2-4} \cmidrule(lr){5-7} \cmidrule(lr){8-10} \cmidrule(lr){11-13}
		h & QS10 & QS50 & QS90 & QS10 & QS50 & QS90 & QS10 & QS50 & QS90 & QS10 & QS50 & QS90 \\
		\midrule
		\multicolumn{13}{@{}l}{\textit{Panel A: 2013M1--2025M2}} \\[0.3ex]
		1  & 1.01 & \textbf{0.99} & 1.22 & 1.31 & 1.00 & 1.24 & 1.11 & \textbf{0.98} & 1.08 & 1.08 & \textbf{0.98} & 1.06 \\
		2  & \textbf{0.98} & \textbf{0.95} & 1.05 & 1.15 & \textbf{0.98} & 1.04 & \textbf{0.97} & \textbf{0.94} & 1.02 & \textbf{0.99} & \textbf{0.96} & \textbf{0.97} \\
		3  & \textbf{0.93} & \textbf{0.93} & 1.02 & 1.16 & \textbf{0.95} & \textbf{0.94} & \textbf{0.95} & \textbf{0.94} & \textbf{0.94} & \textbf{0.99} & \textbf{0.96} & \textbf{0.89} \\
		4  & \textbf{0.93} & \textbf{0.95} & \textbf{0.96} & 1.17 & \textbf{0.96} & \textbf{0.86} & \textbf{0.93} & \textbf{0.95} & \textbf{0.93} & 1.02 & \textbf{0.96} & \textbf{0.89} \\
		5  & \textbf{0.92} & \textbf{0.97} & \textbf{0.97} & 1.14 & \textbf{0.97} & \textbf{0.85} & \textbf{0.96} & \textbf{0.96} & \textbf{0.97} & 1.11 & \textbf{0.96} & \textbf{0.89} \\
		6  & \textbf{0.90} & \textbf{0.96} & 1.02 & 1.10 & \textbf{0.97} & \textbf{0.84} & \textbf{0.95} & \textbf{0.96} & 1.07 & 1.13 & \textbf{0.97} & \textbf{0.96} \\
		9  & \textbf{0.98} & \textbf{0.98} & 1.08 & 1.38 & \textbf{0.97} & \textbf{0.93} & 1.00 & \textbf{0.97} & 1.07 & 1.35 & \textbf{0.96} & 1.11 \\
		12 & \textbf{0.99} & \textbf{0.96} & 1.11 & 1.62 & \textbf{0.98} & \textbf{0.99} & 1.04 & \textbf{0.97} & 1.15 & 1.69 & \textbf{0.96} & 1.18 \\[0.5ex]
		
		\multicolumn{13}{@{}l}{\textit{Panel B: 2008M1--2025M2}} \\[0.3ex]
		1  & \textbf{0.98} & \textbf{0.99} & 1.30 & 1.28 & 1.00 & 1.29 & 1.11 & \textbf{0.99} & 1.12 & 1.08 & \textbf{0.98} & 1.10 \\
		2  & \textbf{0.89} & \textbf{0.97} & 1.12 & 1.11 & \textbf{0.99} & 1.10 & \textbf{0.97} & \textbf{0.96} & 1.07 & 1.00 & \textbf{0.96} & \textbf{0.99} \\
		3  & \textbf{0.86} & \textbf{0.95} & 1.07 & 1.12 & \textbf{0.97} & 1.00 & \textbf{0.96} & \textbf{0.96} & 1.00 & 1.00 & \textbf{0.96} & \textbf{0.92} \\
		4  & \textbf{0.87} & \textbf{0.97} & 1.01 & 1.12 & \textbf{0.97} & \textbf{0.91} & \textbf{0.94} & \textbf{0.95} & \textbf{0.97} & 1.04 & \textbf{0.95} & \textbf{0.90} \\
		5  & \textbf{0.86} & \textbf{0.97} & 1.01 & 1.09 & \textbf{0.98} & \textbf{0.89} & \textbf{0.95} & \textbf{0.95} & \textbf{0.99} & 1.08 & \textbf{0.95} & \textbf{0.90} \\
		6  & \textbf{0.85} & \textbf{0.97} & 1.03 & 1.06 & \textbf{0.97} & \textbf{0.87} & \textbf{0.93} & \textbf{0.94} & 1.08 & 1.09 & \textbf{0.96} & \textbf{0.96} \\
		9  & \textbf{0.89} & \textbf{0.97} & 1.05 & 1.24 & \textbf{0.97} & \textbf{0.92} & \textbf{0.94} & \textbf{0.95} & 1.09 & 1.23 & \textbf{0.94} & 1.10 \\
		12 & \textbf{0.93} & \textbf{0.95} & 1.08 & 1.42 & \textbf{0.98} & \textbf{0.97} & \textbf{0.95} & \textbf{0.97} & 1.19 & 1.57 & \textbf{0.94} & 1.19 \\
		\bottomrule
	\end{tabular}
	
	\vspace{0.5em}
	\begin{minipage}{\textwidth}
		\footnotesize
		\textit{Notes:} Quantile score ratios comparing QBVAR(4) to BVAR(12). Values below 1.00 (bold) indicate superior QBVAR performance. QS10, QS50, QS90 denote the 10th, 50th, and 90th percentile quantile scores. JLN = Jurado-Ludvigson-Ng macroeconomic uncertainty; GPRHT = Caldara-Iacoviello geopolitical threats; GZ spread = Gilchrist-Zakraj\v{s}ek credit spread; GECON = global economic conditions index.
	\end{minipage}
\end{table}

Second, the results reveal striking asymmetries in tail forecasting that vary systematically by predictor type. The uncertainty measure JLN delivers substantial left-tail improvements---up to 15\% at horizon 6---while showing no gains (and occasional deterioration) in the right tail. Conversely, the geopolitical risk index GPRHT excels at right-tail forecasting, with improvements of 14--16\% at horizons 4--6, but performs poorly in the left tail. 
The GZ spread and GECON exhibit more balanced performance at medium horizons, with gains concentrated around the median and one tail, though left-tail performance weakens at longer horizons, particularly for GECON. These asymmetries suggest that different economic fundamentals are informative for different parts of the oil price distribution: uncertainty measures help predict downside risk, while geopolitical indicators capture upside risk.

\subsection{Comparison with the no-change forecast}

Beating a simple random walk (RW) remains one of the most difficult challenges in forecasting commodity returns. Table~\ref{tab:2} presents Quantile Score ratios for the QBVAR(4) relative to this benchmark, using four predictors that illustrate distinct forecasting channels: macroeconomic uncertainty (JLN), global economic conditions (GECON), financial stress (GZ spread), and oil market fundamentals (OECD stocks).\footnote{The Online Supplement reports results for all twelve predictors, four different estimation samples, and QBVAR specifications with lag orders $p\in \{1,2,4,12\}$; the conclusions are qualitatively unchanged.}

\begin{table}[t!]
	\centering
	\small
	\caption{Out-of-sample quantile score ratios relative to random walk}
	\label{tab:2}
	\begin{tabular}{@{}c ccc ccc ccc ccc@{}}
		\toprule
		& \multicolumn{3}{c}{JLN} & \multicolumn{3}{c}{GECON} & \multicolumn{3}{c}{GZ spread} & \multicolumn{3}{c}{OECD stocks} \\
		\cmidrule(lr){2-4} \cmidrule(lr){5-7} \cmidrule(lr){8-10} \cmidrule(lr){11-13}
		h & QS10 & QS50 & QS90 & QS10 & QS50 & QS90 & QS10 & QS50 & QS90 & QS10 & QS50 & QS90 \\
		\midrule
		\multicolumn{13}{@{}l}{\textit{Panel A: 2013M1--2025M2}} \\[0.3ex]
		1  & \textbf{0.81} & \textbf{0.94} & 1.34 & \textbf{0.88} & \textbf{0.94} & 1.13 & \textbf{0.84} & \textbf{0.93} & 1.13 & \textbf{0.91} & \textbf{0.93} & 1.09 \\
		2  & \textbf{0.89} & \textbf{0.94} & 1.15 & \textbf{0.92} & \textbf{0.95} & \textbf{0.97} & \textbf{0.87} & \textbf{0.94} & 1.03 & \textbf{0.94} & \textbf{0.94} & \textbf{0.97} \\
		3  & \textbf{0.87} & 1.00 & 1.21 & \textbf{0.96} & 1.00 & \textbf{0.95} & \textbf{0.91} & 1.00 & 1.04 & \textbf{0.98} & \textbf{0.99} & \textbf{0.99} \\
		4  & \textbf{0.87} & 1.00 & 1.13 & \textbf{0.96} & 1.00 & \textbf{0.99} & \textbf{0.88} & \textbf{0.99} & 1.08 & \textbf{0.99} & 1.00 & 1.02 \\
		5  & \textbf{0.82} & 1.00 & 1.14 & 1.00 & 1.00 & 1.00 & \textbf{0.85} & 1.00 & 1.10 & \textbf{0.96} & 1.00 & 1.08 \\
		6  & \textbf{0.82} & 1.01 & 1.15 & 1.04 & 1.00 & 1.06 & \textbf{0.87} & 1.00 & 1.19 & 1.00 & 1.01 & 1.18 \\
		9  & \textbf{0.86} & 1.00 & 1.19 & 1.25 & 1.00 & 1.19 & \textbf{0.88} & 1.00 & 1.16 & 1.24 & 1.01 & 1.34 \\
		12 & \textbf{0.88} & 1.00 & 1.23 & 1.57 & 1.01 & 1.29 & \textbf{0.93} & 1.00 & 1.25 & 1.45 & 1.01 & 1.49 \\[0.5ex]
		\multicolumn{13}{@{}l}{\textit{Panel B: 2008M1--2025M2}} \\[0.3ex]
		1  & \textbf{0.76} & \textbf{0.92} & 1.37 & \textbf{0.85} & \textbf{0.93} & 1.15 & \textbf{0.82} & \textbf{0.91} & 1.13 & \textbf{0.87} & \textbf{0.92} & 1.08 \\
		2  & \textbf{0.79} & \textbf{0.93} & 1.13 & \textbf{0.89} & \textbf{0.94} & \textbf{0.96} & \textbf{0.85} & \textbf{0.94} & 1.02 & \textbf{0.89} & \textbf{0.94} & \textbf{0.94} \\
		3  & \textbf{0.80} & \textbf{0.98} & 1.13 & \textbf{0.95} & \textbf{0.98} & \textbf{0.95} & \textbf{0.89} & \textbf{0.98} & 1.02 & \textbf{0.95} & \textbf{0.98} & \textbf{0.97} \\
		4  & \textbf{0.82} & \textbf{0.99} & 1.07 & \textbf{0.98} & \textbf{0.99} & \textbf{0.95} & \textbf{0.89} & \textbf{0.99} & 1.02 & \textbf{0.99} & \textbf{0.98} & 1.00 \\
		5  & \textbf{0.80} & \textbf{0.99} & 1.05 & 1.01 & 1.00 & \textbf{0.95} & \textbf{0.89} & \textbf{0.99} & 1.04 & \textbf{0.99} & 1.00 & 1.06 \\
		6  & \textbf{0.80} & 1.00 & 1.05 & 1.04 & 1.00 & 1.00 & \textbf{0.89} & 1.00 & 1.11 & 1.01 & 1.00 & 1.13 \\
		9  & \textbf{0.83} & 1.00 & 1.09 & 1.20 & 1.00 & 1.17 & \textbf{0.89} & 1.00 & 1.16 & 1.14 & 1.00 & 1.37 \\
		12 & \textbf{0.87} & 1.00 & 1.14 & 1.52 & 1.00 & 1.32 & \textbf{0.91} & 1.00 & 1.28 & 1.39 & 1.01 & 1.59 \\
		\bottomrule
	\end{tabular}
	
	\vspace{0.5em}
	\begin{minipage}{\textwidth}
		\footnotesize
		\textit{Notes:} Quantile score ratios comparing QBVAR(4) to random walk. Values below 1.00 (bold) indicate superior QBVAR performance. JLN = Jurado-Ludvigson-Ng macroeconomic uncertainty; GECON = global economic conditions index; GZ spread = Gilchrist-Zakraj\v{s}ek credit spread; OECD stocks = OECD crude oil inventories.
	\end{minipage}
\end{table}	

Three findings stand out. First, the QBVAR achieves meaningful improvements in median forecasts (QS50) at short horizons, with gains of 6--9\% for $h \leq 2$, but these advantages dissipate at longer horizons where ratios converge to unity. This pattern of significant short-run predictability that fades over time is consistent with the well-documented difficulty of beating random walks in commodity markets.

Second, the left tail (QS10) reveals the QBVAR's main value proposition. The uncertainty measure JLN delivers remarkable improvements across all horizons and both evaluation periods, with gains ranging from 12\% to 24\%. Financial conditions captured by the GZ spread show similarly consistent left-tail gains of 9--18\%. These results suggest that uncertainty and credit stress provide substantial advance warning of oil price declines---precisely the downside risk that matters most for risk management.

Third, right-tail forecasting (QS90) proves more challenging but shows horizon-dependent gains. At horizons 2--5, real activity (GECON) and inventory measures (OECD stocks) deliver 3--8\% improvements, though these evaporate at longer horizons. The contrast with left-tail results is striking: while uncertainty predicts downside risk persistently, upside risk remains harder to forecast beyond the near term.

Taken together, these results highlight a fundamental asymmetry in oil price
predictability. Measures of uncertainty and financial stress provide advance
signals of elevated downside risk, even though adverse price realizations
themselves tend to occur abruptly. By contrast, upside risks appear to be driven
by demand- or inventory-related developments that are harder to anticipate beyond
short horizons. From a forecasting perspective, these findings underscore the
value of explicitly modeling the conditional distribution: while mean forecasts
offer limited gains over a random walk, quantile-based approaches deliver
economically meaningful improvements precisely where risk management and policy
concerns are most acute.

\subsection{Comparison with BVAR-SV}

Stochastic volatility is known to enhance forecast accuracy, particularly for tail predictions. Table~\ref{tab:3} presents Quantile Score ratios for the QBVAR(4) relative to a BVAR(12) with stochastic volatility (BVAR-SV), using four predictors that span uncertainty (JLN), financial conditions (GZ spread, EBP), and global activity (GECON).\footnote{The Online Supplement reports results for all twelve predictors, four different estimation samples, and QBVAR specifications with lag orders $p\in \{1,2,4,12\}$; the conclusions are qualitatively unchanged.} Since our quantile-based approach inherently captures heteroskedasticity through quantile-specific coefficients, we incorporate stochastic volatility only in the benchmark.

Three findings emerge. First, the QBVAR achieves consistent, if modest, improvements in median forecasts (QS50), with gains of 1--4\% that persist across horizons and evaluation periods. The consistency is notable: for JLN and GZ spread, every QS50 entry falls below unity.

Second, left-tail forecasting (QS10) reveals the QBVAR's comparative advantage. The uncertainty measure JLN delivers gains of 3--24\%, with improvements in 15 of 16 entries. Financial conditions indicators perform similarly: GZ spread shows gains of up to 21\%, and EBP up to 14\%. These results demonstrate that quantile-specific modeling captures downside risk dynamics that even flexible stochastic volatility specifications miss.

\begin{table}[t!]
	\centering
	\small
	\caption{Out-of-sample quantile score ratios relative to BVAR-SV(12)}
	\label{tab:3}
	\begin{tabular}{@{}c ccc ccc ccc ccc@{}}
		\toprule
		& \multicolumn{3}{c}{JLN} & \multicolumn{3}{c}{GZ spread} & \multicolumn{3}{c}{EBP} & \multicolumn{3}{c}{GECON} \\
		\cmidrule(lr){2-4} \cmidrule(lr){5-7} \cmidrule(lr){8-10} \cmidrule(lr){11-13}
		h & QS10 & QS50 & QS90 & QS10 & QS50 & QS90 & QS10 & QS50 & QS90 & QS10 & QS50 & QS90 \\
		\midrule
		\multicolumn{13}{@{}l}{\textit{Panel A: 2013M1--2025M2}} \\[0.3ex]
		1  & \textbf{0.97} & \textbf{0.97} & 1.49 & 1.03 & \textbf{0.98} & 1.29 & 1.07 & \textbf{0.98} & 1.25 & 1.05 & 1.00 & 1.25 \\
		2  & \textbf{0.86} & \textbf{0.98} & 1.28 & \textbf{0.85} & \textbf{0.97} & 1.15 & \textbf{0.89} & \textbf{0.97} & 1.09 & \textbf{0.87} & \textbf{0.99} & 1.10 \\
		3  & \textbf{0.80} & \textbf{0.97} & 1.35 & \textbf{0.82} & \textbf{0.97} & 1.16 & \textbf{0.86} & \textbf{0.97} & 1.04 & \textbf{0.85} & \textbf{0.97} & 1.09 \\
		4  & \textbf{0.79} & \textbf{0.98} & 1.25 & \textbf{0.80} & \textbf{0.96} & 1.19 & \textbf{0.86} & \textbf{0.97} & 1.05 & \textbf{0.87} & \textbf{0.98} & 1.11 \\
		5  & \textbf{0.77} & \textbf{0.99} & 1.27 & \textbf{0.79} & \textbf{0.97} & 1.24 & \textbf{0.86} & \textbf{0.98} & 1.08 & \textbf{0.93} & \textbf{0.98} & 1.15 \\
		6  & \textbf{0.76} & \textbf{0.98} & 1.24 & \textbf{0.80} & \textbf{0.97} & 1.26 & \textbf{0.90} & \textbf{0.98} & 1.17 & \textbf{0.96} & \textbf{0.98} & 1.15 \\
		9  & \textbf{0.84} & \textbf{0.99} & 1.27 & \textbf{0.85} & 1.00 & 1.24 & \textbf{0.95} & \textbf{0.99} & 1.33 & 1.21 & \textbf{0.98} & 1.28 \\
		12 & \textbf{0.81} & 1.00 & 1.36 & \textbf{0.83} & 1.01 & 1.35 & \textbf{0.99} & 1.00 & 1.48 & 1.43 & \textbf{0.99} & 1.41 \\[0.5ex]
		\multicolumn{13}{@{}l}{\textit{Panel B: 2008M1--2025M2}} \\[0.3ex]
		1  & \textbf{0.96} & \textbf{0.98} & 1.53 & 1.06 & \textbf{0.98} & 1.31 & 1.05 & \textbf{0.99} & 1.31 & 1.09 & \textbf{0.99} & 1.31 \\
		2  & \textbf{0.81} & \textbf{0.98} & 1.29 & \textbf{0.88} & \textbf{0.97} & 1.18 & \textbf{0.90} & \textbf{0.97} & 1.11 & \textbf{0.92} & \textbf{0.98} & 1.12 \\
		3  & \textbf{0.78} & \textbf{0.97} & 1.34 & \textbf{0.86} & \textbf{0.97} & 1.19 & \textbf{0.88} & \textbf{0.96} & 1.08 & \textbf{0.91} & \textbf{0.97} & 1.13 \\
		4  & \textbf{0.78} & \textbf{0.97} & 1.25 & \textbf{0.86} & \textbf{0.96} & 1.20 & \textbf{0.89} & \textbf{0.96} & 1.11 & \textbf{0.93} & \textbf{0.97} & 1.12 \\
		5  & \textbf{0.77} & \textbf{0.97} & 1.23 & \textbf{0.85} & \textbf{0.97} & 1.22 & \textbf{0.90} & \textbf{0.97} & 1.14 & \textbf{0.97} & \textbf{0.97} & 1.15 \\
		6  & \textbf{0.77} & \textbf{0.97} & 1.19 & \textbf{0.85} & \textbf{0.96} & 1.24 & \textbf{0.92} & \textbf{0.96} & 1.23 & \textbf{0.99} & \textbf{0.96} & 1.15 \\
		9  & \textbf{0.80} & \textbf{0.97} & 1.23 & \textbf{0.85} & \textbf{0.97} & 1.29 & \textbf{0.91} & \textbf{0.96} & 1.44 & 1.16 & \textbf{0.96} & 1.31 \\
		12 & \textbf{0.79} & \textbf{0.98} & 1.29 & \textbf{0.82} & \textbf{0.98} & 1.41 & \textbf{0.94} & \textbf{0.97} & 1.65 & 1.39 & \textbf{0.97} & 1.46 \\
		\bottomrule
	\end{tabular}
	
	\vspace{0.5em}
	\begin{minipage}{\textwidth}
		\footnotesize
		\textit{Notes:} Quantile score ratios comparing QBVAR(4) to BVAR-SV(12). Values below 1.00 (bold) indicate superior QBVAR performance. JLN = Jurado-Ludvigson-Ng macroeconomic uncertainty; GZ spread = Gilchrist-Zakraj\v{s}ek credit spread; EBP = excess bond premium; GECON = global economic conditions index.
	\end{minipage}
\end{table}

Third, the QBVAR offers \textit{no} improvement for right-tail forecasting (QS90). Every entry exceeds unity, often substantially (1.10--1.50). This asymmetry has a clear interpretation: oil price collapses build through observable financial stress and uncertainty that the QBVAR exploits, while price spikes arrive as less predictable supply disruptions that stochastic volatility handles better through its flexible variance dynamics. 

For practitioners, these results suggest that quantile-based and volatility-based
approaches are complementary rather than competing: stochastic volatility excels at
capturing unanticipated right-tail price movements, while the QBVAR is particularly
effective at identifying persistent downside risks relevant for 
risk management.

\section{Forecast combinations and performance during major oil events}

The results in Section 3 reveal a striking pattern: the QBVAR excels at left-tail forecasting but struggles with the right tail, particularly against the BVAR-SV benchmark. This asymmetry suggests that the two approaches capture fundamentally different aspects of oil price dynamics---the QBVAR exploits observable predictors of downside risk, while stochastic volatility models better accommodate the sudden variance expansions characteristic of price spikes. A natural question arises: can we combine these complementary strengths to achieve improvements across the entire distribution?

This section investigates forecast combination strategies that blend QBVAR and BVAR-SV predictions, then evaluates whether these gains hold during the major oil price events that matter most for risk management.

\subsection{Forecast combination strategies}

We investigate whether convex combinations of forecasts from QBVAR(4) and BVAR-SV(12), one of the most competitive benchmarks, can enhance predictive performance.\footnote{The online supplement considers different lag-length specifications and sample sizes.} To this end, we evaluate three distinct combination strategies.

First, we employ a fixed-weight scheme, defined as:
\begin{equation}
	\hat{q}^{Comb}_t(\tau,h) = \lambda \hat{q}^{QBV}_t(\tau,h) + (1-\lambda) \hat{q}^{BVSV}_t(\tau,h),
	\label{eq:Combination}
\end{equation}
where $\tau \in \{0.1, 0.5, 0.9\}$ denotes the quantile, $\hat{q}^{Comb}_t(\tau,h)$ is the combined quantile forecast, $\hat{q}^{QBV}_t(\tau,h)$ represents the QBVAR forecast, $\hat{q}^{BVSV}_t(\tau,h)$ denotes the BVAR-SV(12) forecast, and $h$ is the forecasting horizon. Here, $\lambda$ is the fixed weight assigned to the QBVAR forecast, unchanged across each forecasting step, predictor, and horizon.

Second, we adopt a dynamic weighting strategy based on historical performance. At each forecasting step, for each horizon and predictor, we compute the weight assigned to QBVAR, $\lambda_t(\tau,h)$, using the most recent $S$ rolling observations:
\begin{equation}
	\lambda_t(\tau,h) = 1 - \frac{\frac{1}{S} \sum_{j=t-h-S}^{t-h} \rho_\tau( y_{j+h} - \hat{q}^{QBV}_j(\tau,h))}{\frac{1}{S} \sum_{j=t-h-S}^{t-h} \rho_\tau( y_{j+h} - \hat{q}^{BVSV}_j(\tau,h))+\frac{1}{S} \sum_{j=t-h-S}^{t-h} \rho_\tau( y_{j+h} - \hat{q}^{QBV}_j(\tau,h))},
	\label{eq:Combination2}
\end{equation}
where $\lambda_t(\tau,h)$ is the weight for QBVAR at time $t$, quantile $\tau$, and horizon $h$. The right-hand side represents the ratio of quantile scores over the past $S$ observations, assigning greater weight to the model with lower (better) quantile scores, thus prioritizing forecasts with stronger recent performance. For this exercise we set $S=50$ months.

Finally, we implement a strategy that optimizes weights at each forecasting step based on recent performance. Using the latest $S$ observations, we determine the optimal weight for each period, applied to the forecast combination. The weight $\lambda_t(\tau,h)$ varies by quantile $\tau$ and horizon $h$, computed as:
\begin{equation}
	\lambda^*_t(\tau,h) = \argmin_{\lambda \in [0,1]} \frac{1}{S} \sum_{j=t-h-S}^{t-h} \rho_\tau( y_{j+h} - \hat{q}^{Comb}_j(\tau,h)),
	\label{eq:Combination3}
\end{equation}
where $\lambda^*_t(\tau,h)$ minimizes the average quantile score over the prior $S$ observations, optimizing the forecast combination for each step. For this exercise, we set $S=75$ months.\footnote{Results using different values for $S$ for both methods are available upon request.}

\subsection{Optimal combination weights across quantiles}

\begin{figure}[H] 
	\centering
	\includegraphics[width=0.93\textwidth]{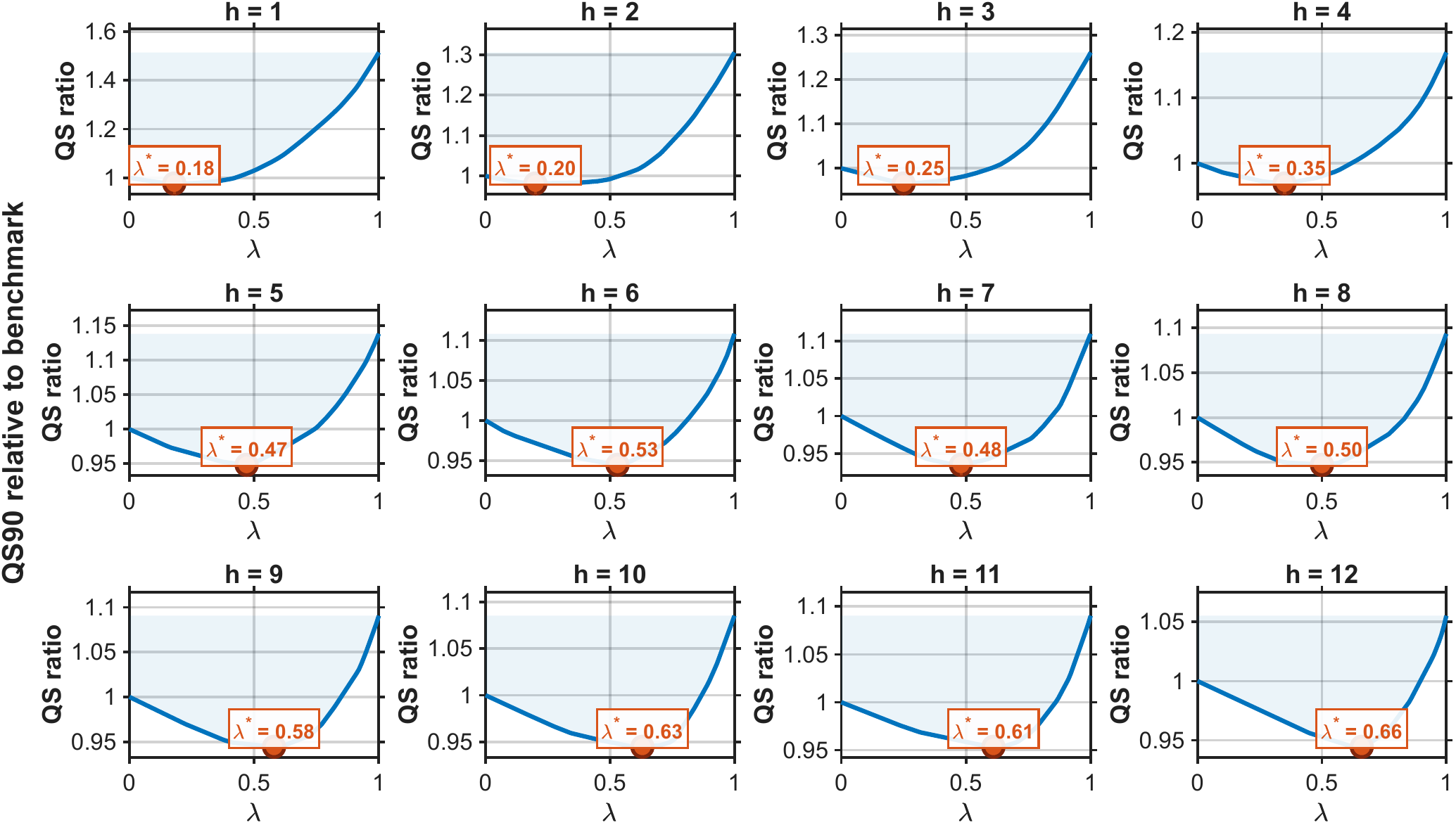} 
	\caption{QS90 of forecast combination relative to BVAR-SV(12), using JLNF12 as predictor. The combined forecast is $\hat{q}_t^{Comb}(\tau,h) = \lambda \hat{q}_t^{QBVAR}(\tau,h) + (1-\lambda) \hat{q}_t^{BVAR-SV}(\tau,h)$, where $\lambda$ is the weight assigned to the QBVAR(4) forecast. $\lambda = 0$ corresponds to pure BVAR-SV(12); $\lambda = 1$ corresponds to pure QBVAR(4). Orange markers indicate the optimal $\lambda^*$ minimizing the quantile score ratio at each horizon.}
	\label{fig:1} 
\end{figure}

\begin{figure}[H] 
	\centering
	\includegraphics[width=0.93\textwidth]{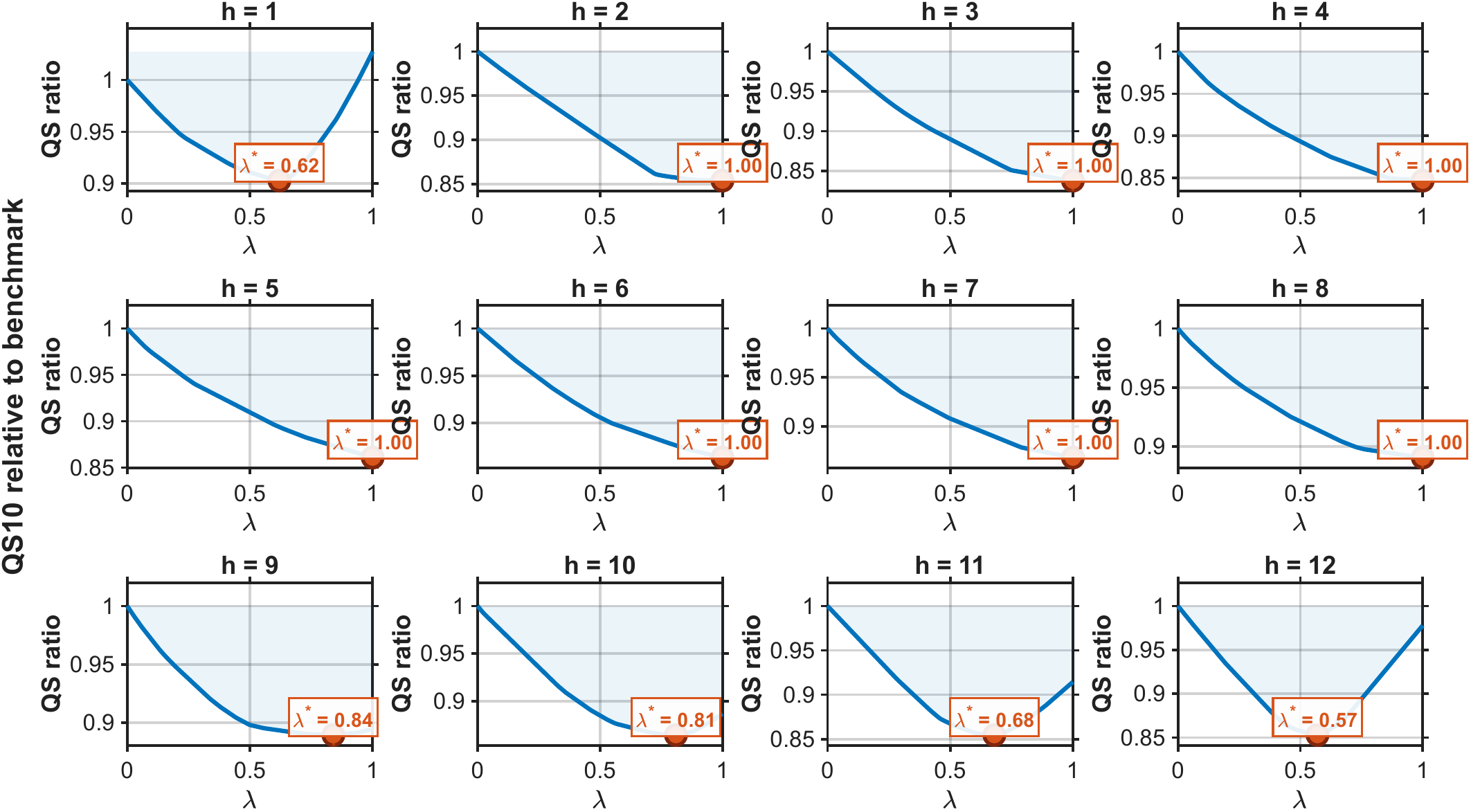} 
	\caption{QS10 of forecast combination relative to BVAR-SV(12), using JLNF12 as predictor. The combined forecast is $\hat{q}_t^{Comb}(\tau,h) = \lambda \hat{q}_t^{QBVAR}(\tau,h) + (1-\lambda) \hat{q}_t^{BVAR-SV}(\tau,h)$, where $\lambda$ is the weight assigned to the QBVAR(4) forecast. $\lambda = 0$ corresponds to pure BVAR-SV(12); $\lambda = 1$ corresponds to pure QBVAR(4). Orange markers indicate the optimal $\lambda^*$ minimizing the quantile score ratio at each horizon.}
	\label{fig:2} 
\end{figure}

To understand when each model contributes most, Figures~\ref{fig:1} and~\ref{fig:2} display Quantile Score ratios for the combined forecast relative to BVAR-SV across a grid of combination weights $\lambda \in [0,1]$, where $\lambda = 1$ corresponds to pure QBVAR and $\lambda = 0$ to pure BVAR-SV. The combined forecast follows equation~(\ref{eq:Combination}) using JLNF12 as the predictor.

For right-tail forecasts (QS90), Figure~\ref{fig:1} shows that intermediate weights
perform best. At horizons $h > 4$, the optimal combination lies near $\lambda = 0.5$,
suggesting QBVAR contains useful right-tail information that is revealed
when combined with BVAR-SV. Notably, $\lambda = 0.4$ improves performance at all
horizons, while $\lambda = 0.5$ does so at all but one.

For left-tail forecasts (QS10), Figure~\ref{fig:2} tells a different story. At 7 of 12 horizons, the optimal weight is $\lambda = 1.0$, meaning QBVAR fully encompasses BVAR-SV for downside risk prediction. Where the optimum falls below unity, it consistently exceeds 0.5, reinforcing the QBVAR's dominance in left-tail forecasting.

These patterns motivate our combination strategies: conservative weights ($\lambda \approx 0.4$--$0.5$) should improve right-tail forecasts while preserving most left-tail gains. Importantly, the weight profiles confirm that QBVAR is the primary
source of improvements in downside risk forecasting, whereas BVAR-SV contributes
mainly to right-tail performance at longer horizons. More broadly, the relative
usefulness of QBVAR and BVAR-SV is quantile- and horizon-dependent, reinforcing
the case for forecast combinations.

\subsection{Fixed-weight combination strategies}

The preceding results highlight a clear complementarity: QBVAR delivers strong left-tail performance, while BVAR-SV contributes mainly to right-tail accuracy at longer horizons. Forecast combinations therefore provide a natural way to exploit these strengths simultaneously. We first examine whether simple averaging rules, applied uniformly across horizons, predictors, and time periods, can outperform BVAR-SV across quantiles. Table~\ref{tab:4} reports Quantile Score ratios for fixed weights $\lambda \in \{0.4, 0.5, 0.6\}$ using two representative predictors: JLNF12 (which showed strong QBVAR performance) and JLN (which showed mixed right-tail results in Section~3).\footnote{Results for additional predictors, sample sizes, and lag-length specifications can be found in the online supplement with similar patterns.}

\begin{table}[t!]
	\centering
	\small
	\caption{Forecast combination with fixed weights: QS ratios relative to BVAR-SV(12)}
	\label{tab:4}
	\resizebox{.8\textwidth}{!}{%
		\begin{tabular}{@{}c ccc ccc ccc@{}}
			\toprule
			& \multicolumn{3}{c}{$\lambda = 0.4$} & \multicolumn{3}{c}{$\lambda = 0.5$} & \multicolumn{3}{c}{$\lambda = 0.6$} \\
			\cmidrule(lr){2-4} \cmidrule(lr){5-7} \cmidrule(lr){8-10}
			h & QS10 & QS50 & QS90 & QS10 & QS50 & QS90 & QS10 & QS50 & QS90 \\
			\midrule
			\multicolumn{10}{@{}c}{\textit{Panel A: JLN}} \\[0.3ex]
			1  & \textbf{0.90} & \textbf{0.99} & 1.00 & \textbf{0.87} & \textbf{0.99} & 1.03 & \textbf{0.86} & \textbf{0.99} & 1.08 \\
			2  & \textbf{0.89} & \textbf{0.99} & \textbf{0.97} & \textbf{0.87} & \textbf{0.99} & \textbf{0.99} & \textbf{0.84} & \textbf{0.99} & 1.02 \\
			3  & \textbf{0.87} & \textbf{0.99} & 1.00 & \textbf{0.85} & \textbf{0.98} & 1.02 & \textbf{0.82} & \textbf{0.98} & 1.05 \\
			4  & \textbf{0.88} & \textbf{0.99} & 1.00 & \textbf{0.86} & \textbf{0.98} & 1.02 & \textbf{0.84} & \textbf{0.98} & 1.04 \\
			5  & \textbf{0.88} & \textbf{0.99} & \textbf{0.99} & \textbf{0.86} & \textbf{0.99} & 1.00 & \textbf{0.84} & \textbf{0.98} & 1.02 \\
			6  & \textbf{0.88} & \textbf{0.99} & \textbf{0.96} & \textbf{0.86} & \textbf{0.98} & \textbf{0.97} & \textbf{0.83} & \textbf{0.98} & \textbf{0.99} \\
			9  & \textbf{0.88} & \textbf{0.99} & \textbf{0.98} & \textbf{0.86} & \textbf{0.98} & \textbf{0.99} & \textbf{0.85} & \textbf{0.98} & 1.00 \\
			12 & \textbf{0.86} & \textbf{0.99} & 1.03 & \textbf{0.83} & \textbf{0.99} & 1.05 & \textbf{0.81} & \textbf{0.99} & 1.07 \\ \hline
			ave & \textbf{0.88} & \textbf{0.99} & \textbf{0.99} & \textbf{0.86} & \textbf{0.99} & 1.01 & \textbf{0.84} & \textbf{0.98} & 1.03 \\  \hline\hline
			\multicolumn{10}{@{}c}{\textit{Panel B: JLNF12}} \\
			1  & \textbf{0.92} & 1.00 & \textbf{0.99} & \textbf{0.91} & \textbf{0.99} & 1.03 & \textbf{0.90} & \textbf{0.99} & 1.08 \\
			2  & \textbf{0.92} & \textbf{0.99} & \textbf{0.98} & \textbf{0.90} & \textbf{0.99} & \textbf{0.99} & \textbf{0.88} & \textbf{0.99} & 1.01 \\
			3  & \textbf{0.91} & \textbf{0.99} & \textbf{0.97} & \textbf{0.89} & \textbf{0.98} & \textbf{0.98} & \textbf{0.87} & \textbf{0.98} & 1.00 \\
			4  & \textbf{0.91} & \textbf{0.98} & \textbf{0.97} & \textbf{0.89} & \textbf{0.98} & \textbf{0.98} & \textbf{0.88} & \textbf{0.98} & 1.00 \\
			5  & \textbf{0.92} & \textbf{0.99} & \textbf{0.95} & \textbf{0.91} & \textbf{0.98} & \textbf{0.95} & \textbf{0.90} & \textbf{0.98} & \textbf{0.97} \\
			6  & \textbf{0.92} & \textbf{0.99} & \textbf{0.95} & \textbf{0.91} & \textbf{0.98} & \textbf{0.95} & \textbf{0.89} & \textbf{0.98} & \textbf{0.95} \\
			9  & \textbf{0.91} & \textbf{0.99} & \textbf{0.95} & \textbf{0.90} & \textbf{0.99} & \textbf{0.95} & \textbf{0.89} & \textbf{0.99} & \textbf{0.94} \\
			12 & \textbf{0.87} & \textbf{0.99} & \textbf{0.96} & \textbf{0.86} & \textbf{0.99} & \textbf{0.95} & \textbf{0.86} & \textbf{0.99} & \textbf{0.95} \\ \hline
			ave & \textbf{0.91} & \textbf{0.99} & \textbf{0.97} & \textbf{0.90} & \textbf{0.99} & \textbf{0.97} & \textbf{0.89} & \textbf{0.98} & \textbf{0.99} \\ \hline\hline
		\end{tabular}
	}
	\begin{tablenotes}
		\small
		\item \textit{Note:} Combined forecast based on eq.(\ref{eq:Combination}) with fixed weight $\lambda$. Values below 1.00 (bold) indicate the combination outperforms BVAR-SV(12). Evaluation period: 2008M1--2025M2.
	\end{tablenotes}
\end{table}

The results confirm that naive averaging succeeds where pure QBVAR failed. With JLNF12, the conservative weight $\lambda = 0.4$ achieves improvements in all 36 entries across quantiles and horizons. On average, this strategy improves left-tail forecasts by 9\%, median forecasts by 1\%, and, crucially, right-tail forecasts by 4\%. Even more aggressive weights ($\lambda = 0.5, 0.6$) deliver gains in nearly all configurations, with only 1--2 exceptions.

The JLN predictor reveals the limits of fixed weights. While left-tail improvements remain substantial (12--16\% on average), right-tail performance becomes mixed at $\lambda \geq 0.5$, with deterioration at longer horizons. This heterogeneity motivates adaptive weighting schemes.

Taken together, the combination results point to a clear asymmetry. QBVAR is the
dominant source of forecast improvements for downside risk, while stochastic
volatility contributes mainly to right-tail performance at longer horizons. Simple
fixed-weight combinations preserve most of the substantial left-tail gains delivered
by QBVAR while improving right-tail forecasts, without requiring real-time estimation
of time-varying weights. These findings highlight the value of forecast combinations
that exploit the complementary strengths of quantile-based and volatility-based
approaches.

\subsection{Adaptive combination strategies}

\subsubsection{Performance-based weights}

While fixed-weight combinations already deliver robust improvements, allowing
combination weights to adapt to recent forecasting performance may yield additional
gains. Hence, rather than fixing weights ex ante, we allow $\lambda_t(\tau,h)$ to vary
at each forecasting step based on recent performance, following
equation~(\ref{eq:Combination2}).

\begin{table}[H]
	\centering
	\footnotesize
	\caption{Forecast combination with performance-based weights: QS ratios relative to BVAR-SV(12)}
	\label{tab:5}
	\setlength{\tabcolsep}{7pt}
	\begin{tabular}{@{}c ccc ccc ccc ccc@{}}
		\toprule
		& \multicolumn{3}{c}{JLNF12} & \multicolumn{3}{c}{GPRHA} & \multicolumn{3}{c}{JLN} & \multicolumn{3}{c}{GZ spread} \\
		\cmidrule(lr){2-4} \cmidrule(lr){5-7} \cmidrule(lr){8-10} \cmidrule(lr){11-13}
		h & QS10 & QS50 & QS90 & QS10 & QS50 & QS90 & QS10 & QS50 & QS90 & QS10 & QS50 & QS90 \\
		\midrule
		1  & \textbf{0.91} & \textbf{0.99} & 1.00 & \textbf{0.99} & 1.00 & 1.00 & \textbf{0.87} & \textbf{0.99} & 1.01 & \textbf{0.92} & \textbf{0.99} & 1.01 \\
		2  & \textbf{0.90} & \textbf{0.99} & \textbf{0.99} & \textbf{0.97} & 1.00 & \textbf{0.99} & \textbf{0.86} & \textbf{0.99} & \textbf{0.98} & \textbf{0.91} & \textbf{0.98} & \textbf{0.99} \\
		3  & \textbf{0.88} & \textbf{0.98} & \textbf{0.98} & \textbf{0.95} & 1.00 & 1.01 & \textbf{0.84} & \textbf{0.98} & 1.01 & \textbf{0.90} & \textbf{0.98} & 1.00 \\
		4  & \textbf{0.89} & \textbf{0.98} & \textbf{0.98} & \textbf{0.97} & \textbf{0.99} & 1.00 & \textbf{0.85} & \textbf{0.98} & 1.02 & \textbf{0.90} & \textbf{0.98} & 1.03 \\
		5  & \textbf{0.91} & \textbf{0.98} & \textbf{0.95} & \textbf{0.97} & 1.00 & \textbf{0.97} & \textbf{0.86} & \textbf{0.99} & 1.00 & \textbf{0.91} & \textbf{0.98} & 1.03 \\
		6  & \textbf{0.91} & \textbf{0.98} & \textbf{0.95} & \textbf{0.97} & 1.00 & \textbf{0.95} & \textbf{0.86} & \textbf{0.98} & \textbf{0.97} & \textbf{0.91} & \textbf{0.98} & 1.03 \\
		9  & \textbf{0.92} & \textbf{0.99} & \textbf{0.95} & \textbf{0.99} & \textbf{0.99} & \textbf{0.94} & \textbf{0.88} & \textbf{0.98} & 1.00 & \textbf{0.90} & \textbf{0.98} & 1.05 \\
		12 & \textbf{0.89} & \textbf{0.99} & \textbf{0.95} & \textbf{0.98} & 1.00 & \textbf{0.97} & \textbf{0.84} & \textbf{0.99} & 1.05 & \textbf{0.87} & \textbf{0.99} & 1.13 \\
		\midrule
		ave & \textbf{0.90} & \textbf{0.99} & \textbf{0.97} & \textbf{0.97} & 1.00 & \textbf{0.98} & \textbf{0.86} & \textbf{0.99} & 1.01 & \textbf{0.91} & \textbf{0.98} & 1.03 \\
		\bottomrule
	\end{tabular}
	\begin{tablenotes}
		\small
		\item \textit{Note:} Combined forecast based on eq.(\ref{eq:Combination2}) with time-varying weights determined by relative historical performance. Values below 1.00 (bold) indicate the combination outperforms BVAR-SV(12). Evaluation period: 2008M1--2025M2.
	\end{tablenotes}
\end{table}

Table~\ref{tab:5} reports results for four representative predictors. Results indicate that performance-based weighting delivers consistent gains across predictors and horizons. With JLNF12, improvements span all quantiles: 9\% for QS10, 1\% for QS50, and 4\% for QS90 on average. The GPRHA predictor achieves similar success with gains across 35 of 36 entries. Even predictors that showed mixed pure-QBVAR performance (JLN, GZ spread) now deliver left-tail improvements of 9--14\% alongside modest median gains, though right-tail results remain mixed for GZ spread at longer horizons.

These gains are consistent with performance-based schemes endogenously assigning
greater weight to QBVAR during periods of heightened downside risk, when tail dynamics
dominate forecast accuracy. Relative to fixed-weight combinations, the benefits of
adaptation are therefore concentrated in the left tail, while median and right-tail
improvements remain more modest.

Overall, allowing combination weights to adapt to recent forecasting performance yields
incremental improvements over fixed-weight schemes, while preserving, and in some cases
strengthening, the dominant role of QBVAR in forecasting downside risk.

\subsubsection{Optimal period-by-period weights}

While performance-based weighting adapts to recent relative accuracy, it still constrains weights to follow a fixed updating rule. We therefore consider a fully flexible approach that directly optimizes the combination weight at each forecasting step. Specifically, this strategy minimizes the quantile score over recent periods, as defined in equation~(\ref{eq:Combination3}). Table~\ref{tab:6} reports results for four predictors spanning different economic channels.

\begin{table}[H]
	\centering
	\footnotesize
	\caption{Forecast combination with optimal weights: QS ratios relative to BVAR-SV(12)}
	\label{tab:6}
	\setlength{\tabcolsep}{6pt}
	\begin{tabular}{@{}c ccc ccc ccc ccc@{}}
		\toprule
		& \multicolumn{3}{c}{JLNF12} & \multicolumn{3}{c}{GZ spread} & \multicolumn{3}{c}{JLN} & \multicolumn{3}{c}{GECON} \\
		\cmidrule(lr){2-4} \cmidrule(lr){5-7} \cmidrule(lr){8-10} \cmidrule(lr){11-13}
		h & QS10 & QS50 & QS90 & QS10 & QS50 & QS90 & QS10 & QS50 & QS90 & QS10 & QS50 & QS90 \\
		\midrule
		1  & \textbf{0.89} & 1.00 & 1.02 & \textbf{0.92} & \textbf{0.97} & 1.08 & \textbf{0.88} & \textbf{0.98} & 1.02 & \textbf{0.94} &1.00  & 1.05  \\
		2  & \textbf{0.86} & \textbf{0.99} & 1.03 & \textbf{0.85} & \textbf{0.98} & 1.00 & \textbf{0.84} & \textbf{0.98} & 1.01 & \textbf{0.91} & \textbf{0.99} & \textbf{0.99} \\
		3  & \textbf{0.81} & \textbf{0.97} & \textbf{0.98} & \textbf{0.81} & \textbf{0.97} & \textbf{0.99} & \textbf{0.80} & \textbf{0.98} & 1.04 & \textbf{0.87} & \textbf{0.98} & 1.00 \\
		4  & \textbf{0.80} & \textbf{0.98} & 1.04 & \textbf{0.80} & \textbf{0.97} & 1.01 & \textbf{0.75} & \textbf{0.99} & 1.08 & \textbf{0.90} & \textbf{0.98} & 1.00 \\
		5  & \textbf{0.78} & \textbf{0.99} & \textbf{0.96} & \textbf{0.78} & \textbf{0.99} & 1.01 & \textbf{0.72} & 1.00 & 1.04 & \textbf{0.95} & \textbf{0.99} & 1.01 \\
		6  & \textbf{0.80} & \textbf{0.99} & \textbf{0.96} & \textbf{0.76} & \textbf{0.99} & 1.02 & \textbf{0.72} & 1.01 & 1.01 & \textbf{0.93} & 1.00 & 1.00 \\
		9  & \textbf{0.80} & 1.02 & \textbf{0.96} & \textbf{0.81} & 1.02 & \textbf{0.99} & \textbf{0.81} & 1.00 & 1.08 & \textbf{0.88} & \textbf{0.99} & 1.00 \\
		12 & \textbf{0.86} & 1.01 & \textbf{0.95} & \textbf{0.81} & 1.01 & 1.00 & \textbf{0.81} & 1.01 & 1.11 & \textbf{0.95} & 1.00 & 1.00 \\
		\midrule
		ave & \textbf{0.83} & \textbf{0.99} & \textbf{0.99} & \textbf{0.82} & \textbf{0.99} & 1.01 & \textbf{0.79} & \textbf{0.99} & 1.05 & \textbf{0.92} & \textbf{0.99} & 1.01 \\
		\bottomrule
	\end{tabular}
	\begin{tablenotes}
		\small
		\item \textit{Note:} Combined forecast based on eq.(\ref{eq:Combination3}) with weights optimized at each period. Values below 1.00 (bold) indicate the combination outperforms BVAR-SV(12). Evaluation period: 2008M1--2025M2.
	\end{tablenotes}
\end{table}

The results highlight the benefits and limits of fully optimized combination weights. This strategy maximizes left-tail gains: JLN achieves average improvements of 21\%,
while GZ spread and JLNF12 deliver 19\% and 18\%, respectively. Median forecasts show
modest but consistent gains of 0--2\%. Right-tail performance is more nuanced:
JLNF12 improves in 9 of 12 horizons (2\% average gain), while predictors such as JLN
exhibit deterioration at longer horizons, suggesting that aggressive optimization
can overfit right-tail dynamics. The GECON index achieves more balanced improvements, with average gains of 8\% for QS10, 1\% for the median, and no gains for QS90.

\subsection{Forecasting performance during major oil events}

A critical question is whether the QBVAR delivers reliable performance during episodes
of large oil price movements, since accurate tail forecasting is particularly valuable
in such environments. To examine this issue, we evaluate the left-tail forecasting
performance, measured by QS10, of the QBVAR(4) and competing benchmarks during three
major oil price declines: the Global Financial Crisis from 2008 to 2009, the oil price
collapse of 2014 to 2015, and the COVID-19 outbreak in 2020. We also assess right-tail
forecasting performance, measured by QS90, during three major oil price surges: the
Gulf War period from 1990 to 1991, the commodity super-cycle of 2007 to 2008, and the
post-COVID surge associated with the Russian invasion of Ukraine from 2021 to 2022.

\begin{figure}[H]
	\centering
	\includegraphics[width=\textwidth]{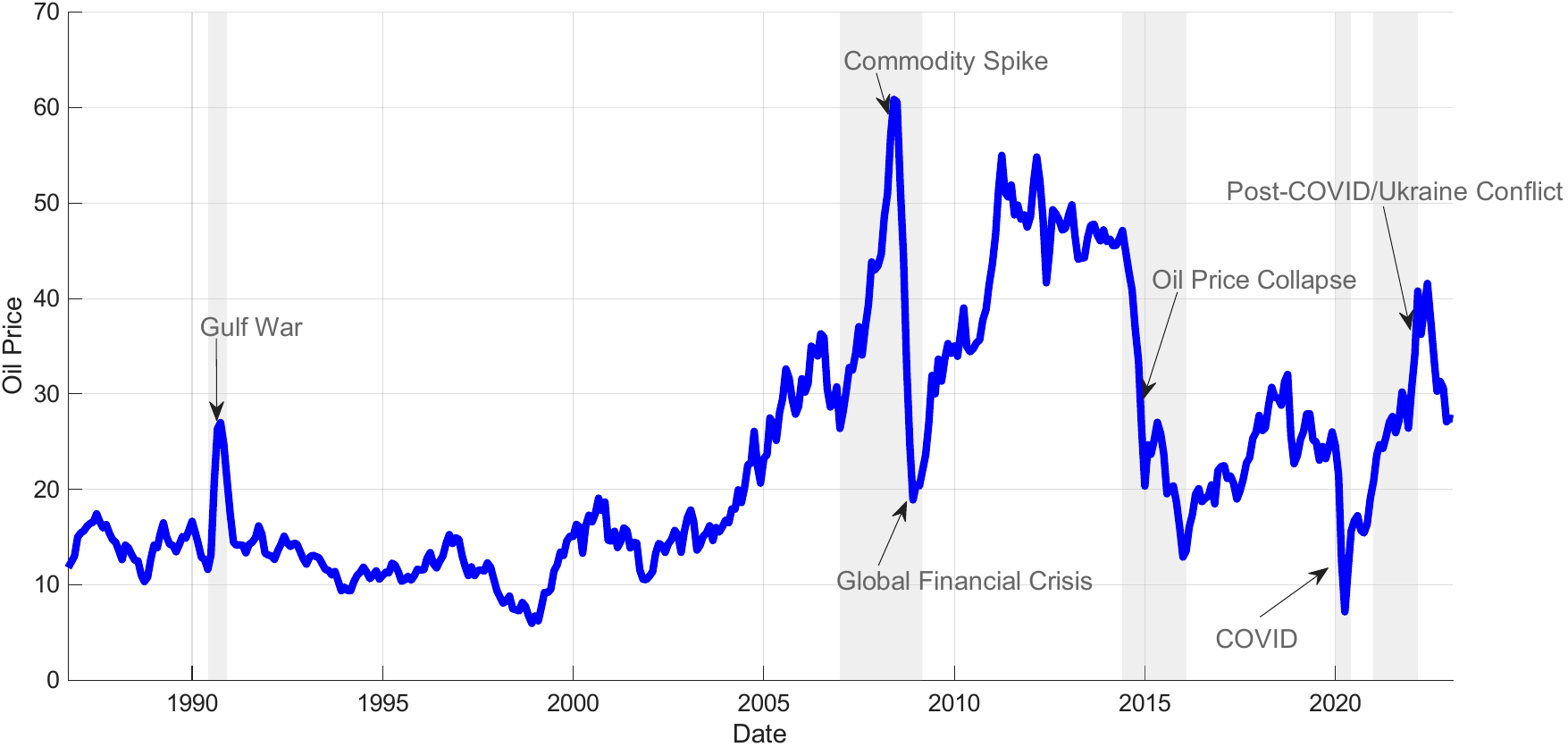}
	\caption{Major oil price events}
	\label{fig:events}
\end{figure}

Figure~\ref{fig:events} illustrates the magnitude of oil price movements during these episodes. In particular, the figure shows that major oil price events are associated with large and rapid movements, often concentrated in the tails of the distribution. This motivates
evaluating forecast performance conditional on such episodes, where tail accuracy is
most economically relevant.

Table~\ref{tab:7} summarizes the key economic and geopolitical mechanisms behind each
episode, distinguishing between negative and positive oil price shocks. This
classification guides the interpretation of tail-specific forecast performance.

\begin{table}[t!]
	\centering
	\footnotesize
	\caption{Major oil price events}
	\label{tab:7}
	\begin{tabular}{lp{5cm}p{4.5cm}p{1.8cm}}
		\toprule
		\textbf{Event} & \textbf{Description} & \textbf{Price Impact} & \textbf{Period} \\
		\midrule
		\multicolumn{4}{l}{\textit{Panel A: Negative Shocks}} \\[0.3ex]
		Financial Crisis & Collapse of financial institutions triggered severe recession & Sharp decline from \$60+ to \$20/bbl & 2008--2009 \\[0.5ex]
		Oil Supply Glut & OPEC maintained production; U.S. shale boom & Decline from \$50+ to sub-\$20/bbl & 2014--2015 \\[0.5ex]
		COVID-19 & Global lockdowns eliminated transportation demand & Fall from \$30+ to sub-\$10/bbl & 2020 \\[0.8ex]
		\multicolumn{4}{l}{\textit{Panel B: Positive Shocks}} \\[0.3ex]
		Gulf War & Iraq's invasion of Kuwait disrupted supply & Price doubled from \$15 to \$28/bbl & 1990--1991 \\[0.5ex]
		Commodity Boom & Strong global growth and emerging market demand & Sustained rise from \$25 to \$60+/bbl & 2007--2008 \\[0.5ex]
		Post-COVID/Ukraine & Economic reopening; Russia-Ukraine conflict & Surge from \$20 to \$40+/bbl & 2021--2022 \\
		\bottomrule
	\end{tabular}
\end{table}

\paragraph{Price Decline Episodes.}

Table~\ref{tab:8} reports QS10 ratios relative to the RW benchmark during price declines, comparing QBVAR against BVAR(12) and BVAR-SV(12). Three patterns emerge. First, RW consistently underperforms. Most entries fall below unity across all models, confirming that structured approaches add value during crises. Second, the QBVAR frequently achieves the largest gains, with improvements reaching 70\% during the 2014--16 collapse (JLNF12 at $h=6$). Third, performance varies by event: all models excel during COVID-19 and the 2014--16 collapse, but results for the 2008 Financial Crisis are more mixed, with BVAR-SV often competitive at short horizons.

\begin{table}[htbp]
	\centering
	\small
	\caption{QS10 ratios relative to RW during price declines}
	\label{tab:8}
	\resizebox{\textwidth}{!}{%
		\begin{tabular}{@{}c ccc ccc ccc ccc@{}}
			\toprule
			& \multicolumn{3}{c}{JLNF12} & \multicolumn{3}{c}{GZ spread} & \multicolumn{3}{c}{JLN} & \multicolumn{3}{c}{EBP} \\
			\cmidrule(lr){2-4} \cmidrule(lr){5-7} \cmidrule(lr){8-10} \cmidrule(lr){11-13}
			h & QVAR & BVAR & BVSV & QVAR & BVAR & BVSV & QVAR & BVAR & BVSV & QVAR & BVAR & BVSV \\
			\midrule
			\multicolumn{13}{@{}l}{\textit{2014--2016 Oil Price Collapse}} \\[0.3ex]
			1  & \textbf{0.99} & \textbf{0.86} & \textbf{0.75} & \textbf{0.89} & \textbf{0.87} & \textbf{0.76} & \textbf{0.93} & \textbf{0.92} & \textbf{0.72} & 1.03 & \textbf{0.82} & \textbf{0.73} \\
			3  & \textbf{0.70} & \textbf{0.88} & \textbf{0.87} & \textbf{0.67} & \textbf{0.85} & \textbf{0.90} & \textbf{0.64} & \textbf{0.83} & \textbf{0.84} & \textbf{0.73} & \textbf{0.86} & \textbf{0.89} \\
			6  & \textbf{0.34} & \textbf{0.68} & \textbf{0.87} & \textbf{0.44} & \textbf{0.65} & \textbf{0.88} & \textbf{0.35} & \textbf{0.69} & \textbf{0.84} & \textbf{0.60} & \textbf{0.66} & \textbf{0.83} \\
			12 & \textbf{0.44} & \textbf{0.63} & \textbf{0.95} & \textbf{0.31} & \textbf{0.64} & \textbf{0.96} & \textbf{0.29} & \textbf{0.61} & \textbf{0.89} & \textbf{0.37} & \textbf{0.68} & \textbf{0.94} \\[0.5ex]
			\multicolumn{13}{@{}l}{\textit{2008 Financial Crisis}} \\[0.3ex]
			1  & 1.01 & 1.02 & \textbf{0.76} & 1.08 & 1.00 & \textbf{0.81} & \textbf{0.75} & 1.01 & \textbf{0.81} & \textbf{0.93} & 1.04 & \textbf{0.81} \\
			3  & \textbf{0.76} & \textbf{0.97} & \textbf{0.91} & \textbf{0.91} & \textbf{0.95} & \textbf{0.88} & \textbf{0.57} & \textbf{0.99} & \textbf{0.96} & \textbf{0.84} & \textbf{0.99} & \textbf{0.94} \\
			6  & \textbf{0.67} & 1.02 & 1.00 & \textbf{0.80} & 1.00 & \textbf{0.95} & \textbf{0.56} & 1.03 & 1.01 & \textbf{0.78} & 1.02 & \textbf{0.97} \\
			12 & \textbf{0.39} & 1.05 & 1.05 & \textbf{0.31} & 1.07 & 1.02 & \textbf{0.33} & 1.06 & 1.10 & \textbf{0.43} & 1.11 & 1.04 \\[0.5ex]
			\multicolumn{13}{@{}l}{\textit{2020 COVID-19 Pandemic}} \\[0.3ex]
			1  & \textbf{0.72} & \textbf{0.85} & \textbf{0.74} & \textbf{0.73} & \textbf{0.71} & \textbf{0.68} & \textbf{0.43} & \textbf{0.65} & \textbf{0.66} & \textbf{0.71} & \textbf{0.82} & \textbf{0.68} \\
			3  & \textbf{0.70} & \textbf{0.87} & \textbf{0.89} & \textbf{0.83} & \textbf{0.90} & \textbf{0.88} & \textbf{0.59} & \textbf{0.82} & \textbf{0.90} & \textbf{0.87} & \textbf{0.91} & \textbf{0.92} \\
			6  & \textbf{0.54} & \textbf{0.86} & \textbf{0.79} & \textbf{0.71} & \textbf{0.89} & \textbf{0.83} & \textbf{0.45} & \textbf{0.85} & \textbf{0.80} & \textbf{0.63} & \textbf{0.92} & \textbf{0.80} \\
			12 & 1.93 & \textbf{0.91} & 1.11 & 1.50 & \textbf{0.87} & 1.17 & 1.92 & 1.24 & 1.12 & 2.15 & \textbf{0.97} & 1.38 \\
			\bottomrule
		\end{tabular}
	}	
	\begin{tablenotes}
		\small
		\item \textit{Note:} QS10 ratios comparing each model to RW during major price decline episodes. Values below 1.00 (bold) indicate superior performance. QVAR = QBVAR(4); BVAR = BVAR(12); BVSV = BVAR-SV(12).
	\end{tablenotes}
\end{table}

\paragraph{Price Surge Episodes.}

Table~\ref{tab:9} reports QS90 ratios during price surges. The patterns differ markedly from declines. First, RW is more competitive---many entries exceed unity, particularly during the 2021--22 crisis. Second, the QBVAR nonetheless achieves substantial gains in specific configurations: during the 2007--08 commodity boom, improvements reach 30\% at medium horizons for several predictors. Third, the Gulf War episode shows the strongest QBVAR performance, with gains of 10--50\% at horizons 9--12 across most predictors. The 2021--22 crisis proves most challenging, with QBVAR underperforming at short horizons but recovering at $h \geq 6$.

\begin{table}[H]
	\centering
	\small
	\caption{QS90 ratios relative to RW during price surges}
	\label{tab:9}
	\resizebox{\textwidth}{!}{%
		\begin{tabular}{@{}c ccc ccc ccc ccc@{}}
			\toprule
			& \multicolumn{3}{c}{OECD stocks} & \multicolumn{3}{c}{EBP} & \multicolumn{3}{c}{World Rigs} & \multicolumn{3}{c}{GECON} \\
			\cmidrule(lr){2-4} \cmidrule(lr){5-7} \cmidrule(lr){8-10} \cmidrule(lr){11-13}
			h & QVAR & BVAR & BVSV & QVAR & BVAR & BVSV & QVAR & BVAR & BVSV & QVAR & BVAR & BVSV \\
			\midrule
			\multicolumn{13}{@{}l}{\textit{2021--2022 Post-COVID/Ukraine Crisis}} \\[0.3ex]
			1  & 1.21 & \textbf{0.97} & 1.14 & 1.38 & \textbf{0.88} & 1.10 & 1.34 & 1.02 & 1.10 & 1.31 & \textbf{0.95} & 1.08 \\
			3  & \textbf{0.85} & 1.22 & 1.23 & \textbf{0.94} & 1.16 & 1.18 & \textbf{0.98} & 1.14 & 1.20 & 1.03 & 1.17 & 1.25 \\
			6  & \textbf{0.72} & 1.03 & 1.08 & \textbf{0.86} & 1.23 & 1.15 & \textbf{0.78} & 1.06 & 1.14 & \textbf{0.78} & 1.06 & 1.07 \\
			12 & 1.16 & 1.09 & 1.25 & 1.01 & 1.21 & 1.32 & \textbf{0.88} & \textbf{0.92} & 1.24 & 1.05 & 1.11 & 1.20 \\[0.5ex]
			
			\multicolumn{13}{@{}l}{\textit{2007--2008 Commodity Super-Cycle}} \\[0.3ex]
			
			1  & \textbf{0.82} & \textbf{0.65} & \textbf{0.78} & \textbf{0.94} & \textbf{0.67} & \textbf{0.82} & 1.10 & \textbf{0.64} & \textbf{0.86} & 1.19 & \textbf{0.67} & \textbf{0.80} \\
			3  & \textbf{0.76} & \textbf{0.65} & \textbf{0.75} & \textbf{0.69} & \textbf{0.68} & \textbf{0.81} & \textbf{0.86} & \textbf{0.70} & \textbf{0.80} & 1.00 & \textbf{0.70} & \textbf{0.78} \\
			6  & \textbf{0.93} & \textbf{0.85} & \textbf{0.81} & \textbf{0.75} & \textbf{0.77} & \textbf{0.91} & \textbf{0.77} & \textbf{0.76} & \textbf{0.76} & \textbf{0.77} & \textbf{0.73} & \textbf{0.84} \\
			12 & 1.17 & \textbf{0.97} & \textbf{0.90} & \textbf{0.95} & 1.05 & \textbf{0.90} & \textbf{0.81} & \textbf{0.95} & \textbf{0.84} & \textbf{0.82} & \textbf{0.91} & \textbf{0.84} \\[0.5ex]
			
			\multicolumn{13}{@{}l}{\textit{1990--1991 Gulf War}} \\[0.3ex]
			1  & \textbf{0.85} & \textbf{0.84} & \textbf{0.89} & \textbf{0.94} & \textbf{0.86} & \textbf{0.98} & \textbf{0.92} & \textbf{0.85} & \textbf{0.94} & \textbf{0.95} & \textbf{0.84} & \textbf{0.89} \\
			3  & \textbf{0.95} & \textbf{0.98} & 1.00 & 1.01 & \textbf{0.97} & 1.03 & \textbf{0.98} & \textbf{0.98} & 1.01 & 1.10 & 1.02 & 1.05 \\
			6  & \textbf{0.92} & 1.08 & 1.02 & \textbf{0.99} & 1.23 & 1.17 & 1.00 & 1.20 & 1.23 & 1.18 & 1.18 & 1.17 \\
			12 & \textbf{0.88} & \textbf{0.52} & \textbf{0.86} & \textbf{0.71} & \textbf{0.54} & \textbf{0.87} & \textbf{0.76} & \textbf{0.51} & \textbf{0.89} & \textbf{0.62} & \textbf{0.50} & \textbf{0.84} \\
			\bottomrule
		\end{tabular}
	}
	\begin{tablenotes}
		\small
		\item \textit{Note:} QS90 ratios comparing each model to RW during major price surge episodes. Values below 1.00 (bold) indicate superior performance. QVAR = QBVAR(4); BVAR = BVAR(12); BVSV = BVAR-SV(12).
	\end{tablenotes}
\end{table}

Taken together, the event-based analysis confirms and sharpens the full-sample
findings. During major oil price declines, the QBVAR consistently delivers the
largest improvements in left-tail forecasting accuracy, particularly at medium and
longer horizons, highlighting the importance of quantile-specific dynamics when
financial stress and uncertainty intensify. By contrast, performance during price
surges is more heterogeneous, with QBVAR gains concentrated in specific episodes
such as the Gulf War and the commodity super-cycle, and weaker results during the
post-COVID period. These results underscore a fundamental asymmetry in oil price
dynamics: downside risks tend to build gradually through observable macro-financial
channels, while upside risks are more episodic and driven by hard-to-predict supply
disruptions.

\section{Conclusions}\label{sec:Conclusion}

This paper develops a Bayesian Quantile Vector Autoregression that models the entire conditional distribution of oil prices through quantile-specific coefficients and factor structures. Our out-of-sample evaluation spanning 1975--2025 yields three main findings. First, the QBVAR improves median forecasts by 2--5\% relative to standard Bayesian VARs, demonstrating that quantile-specific dynamics benefit even point prediction. Second, uncertainty and financial condition variables strongly predict downside risk, with left-tail improvements of 10--25\% that are particularly pronounced during crisis episodes. Third, right-tail forecasting remains challenging: the QBVAR typically underperforms stochastic volatility models for the 90th percentile, suggesting oil price collapses build through observable financial stress while spikes arrive as less predictable supply disruptions.

These findings have practical implications for risk assessment. Traditional models focusing on expected price movements may systematically underestimate downside risks during periods of rising financial stress or macroeconomic uncertainty. Our results show that financial conditions and uncertainty measures substantially improve left-tail predictions, providing earlier warning of potential price collapses. This is valuable for investors managing portfolio risk, firms hedging commodity exposure, and policymakers concerned with macroeconomic stability. The persistent difficulty in forecasting right-tail events suggests that upside risks may require hybrid approaches combining quantile-specific coefficients with stochastic volatility, or alternative predictors designed to capture supply disruption risks.

\newpage
\begin{appendix}
	\renewcommand{\theequation}{A.\arabic{equation}} \setcounter{equation}{0} %
	\renewcommand{\thetable}{A\arabic{table}} \setcounter{table}{0}
	\renewcommand{\thefigure}{A\arabic{figure}} \setcounter{figure}{0}
	\section{Data Appendix}
	\label{sec:dataAppendix}
	
	Our analysis employs monthly data spanning from January 1975 to February 2025, encompassing oil price series and a comprehensive set of predictor variables. All transformations follow standard practices in the oil price forecasting literature, with transformation codes (tcode) indicating the specific data treatment applied to achieve stationarity and comparability across variables.
	
	\paragraph{Oil Price Data} We construct two primary oil price series, both expressed in real terms using the U.S. Consumer Price Index for All Urban Consumers (CPIAUCSL) as the deflator.
	
	\textbf{Refiner Acquisition Cost (RAC):} This series represents the refiner acquisition cost of imported crude oil, obtained from Table 9.1 of the Energy Information Administration's Monthly Energy Review (MER). The RAC provides a comprehensive measure of crude oil costs faced by U.S. refiners and serves as a reliable proxy for oil price movements affecting the domestic market.
	
	\textbf{Brent Crude Oil (BRENT):} We construct a monthly Brent crude oil price series using EIA spot price data available from May 1987 onwards. To extend this series backward to January 1975, we employ the growth rate of the monthly refiner acquisition cost of imported crude oil from the MER to extrapolate Brent prices. This backcasting methodology ensures consistency with historical price movements while providing the longest possible time series for analysis.
	
	Both oil price series undergo logarithmic transformation (tcode = 5) to achieve stationarity and facilitate interpretation of results in terms of percentage changes.
	
	\paragraph{Predictor Variables} Our forecasting models incorporate a diverse set of predictor variables capturing different dimensions of oil market fundamentals and broader economic conditions.
	
	\textbf{Global Economic Activity Measures:} The Global Economic Conditions Index (GECON) from \cite{BaumeisterKorobilisLee2022} aggregates 16 components to provide a comprehensive measure of worldwide economic activity (tcode = 1, indicating first differences). The Kilian Index offers an alternative measure based on dry cargo bulk freight volumes adjusted for shipping costs, reflecting global demand for industrial inputs that are not readily storable \citep{Kilian2009}.
	
	\textbf{Supply-Side Indicators:} Worldwide Oil Rig Counts from Baker Hughes serve as a forward-looking indicator of oil production capacity, as drilling activity responds to anticipated future oil prices. A decline in rig counts typically signals reduced future supply and upward pressure on prices. Global crude oil production data from the EIA provides a direct measure of current supply conditions.
	
	\textbf{Inventory and Storage:} OECD crude oil inventories are constructed following \cite{Hamilton2009} and \cite{KilianMurphy2014} methodologies. Due to limited availability of crude oil stock data for non-U.S. OECD countries, we adjust U.S. crude oil stockpiles by the ratio of OECD petroleum reserves to U.S. petroleum reserves. For the period prior to January 1988, when reliable OECD petroleum stock data are unavailable, we backcast using U.S. petroleum inventory growth rates.
	
	\textbf{Consumption and Demand:} Global Liquid Fuels Consumption (GLFC) data from the EIA's Short-Term Energy Outlook provides monthly consumption figures in million barrels per day starting January 1990. We extend this series backward to January 1982 using OECD refined petroleum products consumption growth rates, and further to January 1975 using world crude oil production growth rates from the International Energy Statistics.
	
	\textbf{Geopolitical Risk:} Three variants of the Geopolitical Risk Index (GPRH, GPRHT, GPRHA) from \cite{CaldaraIacoviello2022} capture different aspects of global political risk through automated text analysis of news coverage. These indices quantify threats, realizations, and escalations of adverse geopolitical events including wars, terrorism, and international tensions.
	
	\textbf{Financial Market Conditions:} The Excess Bond Premium (EBP) and GZ Corporate Spread from \cite{GilchristZakrajsek2012} measure corporate credit market conditions and financial stress. The EBP represents the component of corporate bond spreads unexplained by default risk, reflecting investor risk appetite, market liquidity, and financial intermediary balance sheet health.
	
	\textbf{Product Market Spreads:} The crack spread measures the differential between gasoline and crude oil prices, calculated using unleaded regular gasoline prices from the MER Table 9.4. Gasoline prices are converted from dollars per gallon to dollars per barrel and compared with Brent prices. For the period before January 1976, we backcast using leaded gasoline price growth rates.
	
	\textbf{Uncertainty Measures:} Macroeconomic (JLN12) and Financial (JLNF12) uncertainty indices from \cite{jurado2015measuring} provide model-based measures of uncertainty extracted from the conditional volatility of multiple economic and financial time series. These indices offer more comprehensive uncertainty measures than traditional single-variable indicators like the VIX.
	
	\paragraph{Data Transformations}
	
	All variables undergo appropriate transformations to ensure stationarity and comparability. Transformation codes indicate: tcode = 1 for first differences (levels to growth rates), tcode = 5 for logarithmic first differences (levels to log growth rates). Variables already in growth rate or index form (tcode = 1) enter the analysis in first differences, while level variables expressing quantities or prices (tcode = 5) are log-differenced to obtain growth rates.
	
	\paragraph{Data Sources and Availability}
	
	Primary data sources include the Energy Information Administration (EIA)\footnote{See \href{https://www.eia.gov/totalenergy/data/monthly/}{https://www.eia.gov/totalenergy/data/monthly/}.} for oil market data, the Federal Reserve Economic Data (FRED) database for macroeconomic indicators, and specialized academic datasets for uncertainty and geopolitical risk measures. The sample period begins in January 1975 to maximize the time series length while ensuring data quality and consistency across variables. Where necessary, careful backcasting procedures extend key series using correlated indicators with longer historical availability. Table~\ref{tab:data_description} provides a comprehensive summary of all variables, their sources, and transformation codes.
	
	\begin{table}[H]
		\centering
		\small
		\caption{Data Description and Sources}
		\label{tab:data_description}
		\begin{tabular}{@{}llcc@{}}
			\toprule
			\textbf{Variable} & \textbf{Description} & \textbf{Source} & \textbf{tcode} \\
			\midrule
			\multicolumn{4}{@{}l}{\textit{Oil Prices}} \\[0.3ex]
			RAC & Refiner Acquisition Cost (real) & EIA$^a$ & 5 \\
			BRENT & Brent spot price (real) & EIA$^a$ & 5 \\[0.5ex]
			\multicolumn{4}{@{}l}{\textit{Global Activity}} \\[0.3ex]
			GECON & Global Economic Conditions Index & [1] & 1 \\
			Kilian Index & Real Economic Activity (shipping) & FRED & 1 \\
			GLFC & Global Liquid Fuels Consumption & EIA$^b$ & 5 \\[0.5ex]
			\multicolumn{4}{@{}l}{\textit{Supply Indicators}} \\[0.3ex]
			Oil Rigs & Worldwide Oil Rig Counts & Baker Hughes & 5 \\
			Global crude oil & Global crude oil production & EIA & 5 \\
			OECD stocks & OECD crude oil inventories & EIA$^c$ & 5 \\[0.5ex]
			\multicolumn{4}{@{}l}{\textit{Geopolitical Risk}} \\[0.3ex]
			GPRH & Geopolitical Risk Index & [2] & 1 \\
			GPRHT & Geopolitical Threats Index & [2] & 2 \\
			GPRHA & Geopolitical Acts Index & [2] & 2 \\[0.5ex]
			\multicolumn{4}{@{}l}{\textit{Financial Conditions}} \\[0.3ex]
			EBP & Excess Bond Premium & [3] & 1 \\
			GZ spread & GZ Corporate Spread & [3] & 5 \\
			Crack spread & Gasoline--Brent differential & EIA$^a$ & 1 \\[0.5ex]
			\multicolumn{4}{@{}l}{\textit{Uncertainty}} \\[0.3ex]
			JLN & Macroeconomic Uncertainty (12-month) & [4] & 2 \\
			JLNF12 & Financial Uncertainty (12-month) & [4] & 1 \\[0.5ex]
			\multicolumn{4}{@{}l}{\textit{Auxiliary}} \\[0.3ex]
			CPIAUCSL & U.S. Consumer Price Index & FRED & -- \\
			OECD consumption & OECD Petroleum Products Consumption & EIA$^c$ & -- \\
			\bottomrule
		\end{tabular}
		
		\vspace{0.5em}
		\begin{minipage}{\textwidth}
			\footnotesize
			\textit{Notes:} Transformation codes: tcode = 1 (first difference), tcode = 2 (no transformation), tcode = 5 (log first difference). All oil prices deflated by CPI. Sample: 1975:M1--2025:M2.\\[0.3em]
			\textit{Sources:} $^a$Monthly Energy Review; $^b$Short-Term Energy Outlook; $^c$International Energy Statistics.\\[0.3em]
			\textit{References:} [1]~\cite{BaumeisterKorobilisLee2022}; [2]~\cite{CaldaraIacoviello2022}; [3]~\cite{GilchristZakrajsek2012}; [4]~\cite{jurado2015measuring}.
		\end{minipage}
	\end{table}

\end{appendix}

\bibliographystyle{apalike}
\addcontentsline{toc}{section}{\refname}
\bibliography{OILATRISK.bib}

\end{document}